\documentclass[journal]{IEEEtran}
\usepackage{cite}
\usepackage{amsmath,amssymb,amsfonts,amsthm}

\usepackage[utf8]{inputenc} 
\usepackage[T1]{fontenc}

\usepackage{graphicx}
\usepackage{textcomp}
\usepackage{xcolor,soul}
\usepackage{stmaryrd}
\usepackage{algpseudocode}
\usepackage{amsmath}
\usepackage{mathrsfs}
\usepackage{bm}
\usepackage[abs]{overpic}
\usepackage[nocomma]{optidef}

\usepackage{physics}
\usepackage{tikz}
\usepackage{mathdots}
\usepackage{yhmath}
\usepackage{cancel}
\usepackage{siunitx}
\usepackage{array}
\usepackage{multirow}
\usepackage{amssymb}
\usepackage{gensymb}
\usepackage{tabularx}
\usepackage{extarrows}
\usepackage{booktabs}
\usepackage[ruled,linesnumbered]{algorithm2e}
\makeatletter
\newcommand{\mathleft}{\@fleqntrue\@mathmargin0pt}
\newcommand{\mathcenter}{\@fleqnfalse}
\makeatother

\usetikzlibrary{fadings}
\usetikzlibrary{patterns}
\usetikzlibrary{shadows.blur}
\usetikzlibrary{shapes}

\theoremstyle{plain}
\newtheorem*{theorem*}{Theorem}

\newtheorem*{eg}{Example}

\usepackage{xcolor,varwidth}

\usepackage[scaled=.75]{beramono}
\newcommand{\bpara}[1]		{\medskip \noindent {\bf #1}}

\renewcommand\geq\geqslant
\renewcommand\leq\leqslant

\newcommand\fig[1]			{Fig.~\ref{#1}}

\def\eg					    {\emph{e.g.,}~}
\def\ie					    {\emph{i.e.,}~}

\usepackage{xspace}

\DeclareDocumentCommand{\sd}{o}  
{{\underline{\ast}\IfValueT{#1}{_{#1}}}}

\DeclareDocumentCommand{\xmod}{m o o o}  
{%
\IfNoValueTF{#4}
{{#1}\IfValueT{#2}{_{m,\mathsf{#2}}}\IfValueT{#3}{#3}}
{{#1}\IfValueT{#2}{_{m,\mathsf{#2}}^{\mathsf{#4}}}\IfValueT{#3}{#3}}
}

\usepackage[inline]{enumitem}
\DeclareMathAlphabet{\mathsfit}{T1}{\sfdefault}{\mddefault}{\sldefault}
\SetMathAlphabet{\mathsfit}{bold}{T1}{\sfdefault}{\bfdefault}{\sldefault}
\usepackage[switch]{lineno}
\usepackage{siunitx}
\def\BibTeX{{\mathrm B\kern-.05em{\sc i\kern-.025em b}\kern-.08em
    T\kern-.1667em\lower.7ex\hbox{E}\kern-.125emX}}
    
\usepackage{hyperref}
\usepackage{etoolbox}

\begin{document}

%\title{Max-Min Fairness Design for Energy-Efficient Quantized ISAC LEO Satellite Systems: \\ A Rate-Splitting Approach
\title{Non-reciprocal Beyond Diagonal RIS: Sum-Rate Maximization in Full-Duplex Communications
% Max-Min Fair Energy-Efficient Beam Design\\ for Quantized ISAC LEO Satellite Systems:\\ A Rate-Splitting Approach
% \thanks{This work has been partially supported by UKRI grant EP/X040569/1, EP/Y037197/1, EP/X04047X/1, EP/Y037243/1.}
}

\author{
Ziang Liu,
        Hongyu Li,
        and~Bruno Clerckx,~\IEEEmembership{Fellow,~IEEE}
\thanks{Z. Liu, H. Li  and B. Clerckx are with the Communications \& Signal Processing (CSP) Group at the Dept. of Electrical and Electronic Engg., Imperial College London, SW7 2AZ, UK. (e-mails:\{ziang.liu20, c.li21, b.clerckx\}@imperial.ac.uk).
}}
% <-this % stops a space
% \thanks{J. Doe and J. Doe are with Anonymous University.}% <-this % stops a space

% \and
% \and
% \and
% \and
% \and

\maketitle

\begin{abstract}
Reconfigurable intelligent surface (RIS) has been envisioned as a key technology in future wireless communication networks to enable smart radio environment. To further enhance the passive beamforming capability of RIS, beyond diagonal (BD)-RIS has been proposed considering reconfigurable interconnections among different RIS elements. BD-RIS has a unique feature that cannot be enabled by conventional diagonal RIS (D-RIS); it can be realized by non-reciprocal circuits and thus enables an asymmetric scattering matrix. This feature provides the capability to break the wireless channel reciprocity, and has the potential to benefit full-duplex (FD) systems. 
{In this paper, we model the BD RIS-assisted FD systems, where the impact of BD-RIS non-reciprocity and that of structural scattering, which refers to the specular reflection generated by RIS when the RIS is turned OFF, are explicitly captured.} To assess the benefits of non-reciprocal BD-RIS, we optimise the scattering matrix, precoder and combiner to maximize the DL and UL sum-rates in the FD system. To tackle this optimization problem, we propose an iterative algorithm based on block coordination descent (BCD) and penalty dual decomposition (PDD). Numerical results demonstrate surprising benefits of non-reciprocal BD-RIS that it can achieve much higher DL and UL sum-rates in the FD scenario than reciprocal BD-RIS and conventional D-RIS.
\end{abstract}


\section{Introduction}
Reconfigurable intelligent surface (RIS) has been recognized as a key technology to enable beyond 5G and 6G wireless network due to its potential to manipulate the wireless propagation channel and enhance the spectrum and energy efficiency using low-cost hardware \cite{di_renzo_smart_2020, basar2019wireless, bjornson2020reconfigurable}. RIS is a planar surface composed of passive reconfigurable scattering elements, whose phase shift can be tuned to facilitate passive beamforming for both incident and reflected directions. Consequently, the incident, reflected, refracted, and scattered signals from RISs can be intentionally controlled without amplification, allowing manipulation of the radio signal's propagation environment and enhancing the received signal power at the receiver. Thanks to these appealing features, RIS has been extensively studied to improve sum-rates \cite{guo2020weighted} and energy efficiency \cite{huang2019reconfigurable} of the wireless communication systems. In addition, RIS has demonstrated effectiveness in assisting various wireless applications, such as assisting full-duplex (FD) systems \cite{sharma_intelligent_2021, tota_khel_performance_2022}, enabling integrated sensing and communications (ISAC) \cite{chen2023simultaneous}, and interplaying with rate-splitting multiple access (RSMA) to enhance performance \cite{niu2023active}.

% In addition, RISs inherently support full-duplex (FD) transmission due to the channel reciprocity and no noise corruption \cite{basar2019wireless}. These appealing features have motivated research on RIS-assisted in-band FD, where the IBFD operation enables the communication transceiver to transmit and receive simultaneously on the same frequency band \cite{sabharwal_band_2014}. In FD case, the challenge lies on the signal of interest (SoI) drowning in the self interference (SI), due the coupling between transmitter and receiver. Therefore, many research has been conducted on boosting the SoI, and thus reduce the effect of SI via RISs \cite{sharma_intelligent_2021, tota_khel_performance_2022}.

% The RISs have a number of benefits and gained extensive attention. Specifically, it is (1) nearly passive and thus requires no power supply and components in RF chain, including digital to analog converter (DAC), analog-to-digital converter (ADC), and low-noise amplifier (LNA), (2) it will not induce noise, (3) it has response on any frequency \cite{basar2019wireless}. It has 

From the technical perspective, RIS can be modeled as multiple scattering elements connected to a reconfigurable impedance network \cite{shen_modeling_2022, li_reconfigurable_2024, pozar_microwave_2021}. All the above research \cite{guo2020weighted, huang2019reconfigurable,sharma_intelligent_2021, tota_khel_performance_2022,chen2023simultaneous,niu2023active} focuses on conventional RIS with a single-connected architecture, also called Diagonal RIS (D-RIS), in which each port of the reconfigurable impedance network is connected to its own impedance component to ground. This simple architecture allows only the phase of the impinging waves to be tuned and thus limits the capability to manipulate the propagation environment \cite{shen_modeling_2022}.  To enhance the functionality of RISs, RIS 2.0, known as Beyond Diagonal RIS (BD-RIS), has been proposed in \cite{shen_modeling_2022, li_reconfigurable_2024}. In this architecture, ports are inter-connected, enabling the impinging waves to be reconfigured as they flow through the surfaces. Consequently both the phases and magnitudes of the impinging waves can be adjusted. This enhances the ability of controlling passive beamforming of the RIS, resulting in a further improvement of the received signal power at the expense of higher circuit complexity. In the family of BD-RIS, fully-connected architecture has been first proposed, in which all ports are connected to each other via impedance components, leading to a full scattering matrix \cite{shen_modeling_2022}. 
To reduce the required number of impedance components while keeping the wave manipulation flexibility, group-connected architecture with a block-diagonal scattering matrix has been proposed \cite{shen_modeling_2022}, in which ports are divided into groups with each group being fully-connected. To show the benefits of group/fully-connected BD-RIS, a closed-form optimization has been proposed in \cite{nerini2023closed} to maximize the received power. 
In addition, to find the best performance-circuit complexity trade-off, tree- and forest-connected architectures have been proposed and found to achieve the performance-complexity Pareto frontier \cite{nerini2024beyond, nerini2023pareto}.

\begin{figure}
    \centering
    \includegraphics[width=0.45\linewidth]{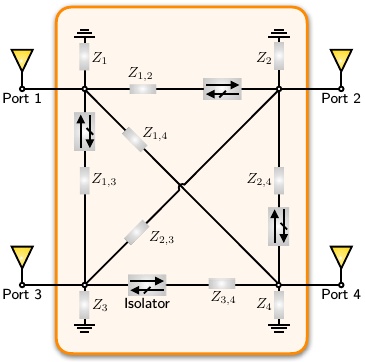}
    \caption{{Hardware implementation of 4-port non-reciprocal BD-RIS.}}
    \label{fig:nr_hardware}
\end{figure}

Based on the reciprocity of the impedance network, the BD-RIS can be further classified into reciprocal and non-reciprocal BD-RIS, \ie the reciprocal BD-RIS has symmetric scattering matrix and non-reciprocal BD-RIS has asymmetric scattering matrix. The reciprocal one is widely studied in existing research \cite{di_renzo_smart_2020, shen_modeling_2022, li_reconfigurable_2024, li2022beyond, nerini2024beyond}, realized by reciprocal circuits with various circuit topologies. A representative example is the group and fully-connected architectures proposed in \cite{shen_modeling_2022}. However, non-reciprocal BD-RIS has received limited investigation in the RIS literature. The non-reciprocal impedance network in the non-reciprocal BD-RIS relaxes the symmetry constraint, resulting in potential flexibility to improve system performances and enable new applications. {In \cite{li_reconfigurable_2022}, a novel architecture is proposed for non-diagonal BD-RIS based on phase shifting. Additionally, \cite{zhang2018space, zhang2019breaking} demonstrate that reciprocity can be broken by properly customizing the coding sequence in field-programmable gate arrays (FPGAs) integrated with RIS. Non-reciprocity can also be achieved by adding non-reciprocal microwave devices, \eg isolator and circulator \cite{pozar_microwave_2021} in the BD-RIS impedance networks. An example of 4-port non-reciprocal BD-RIS hardware implementation by such methods is given in \fig{fig:nr_hardware} \cite{pozar_microwave_2021, xu2024non}.} These works provide a foundation for leveraging non-reciprocal BD-RIS in wireless communications. More potential applications have been envisioned in \cite{pan_full_duplex_2021}, such as utilizing non-reciprocal BD-RIS to enable secure information transmission by introducing interfering signals in specific directions and delivering energy to designated targets. In \cite{wang2024channel}, channel reciprocity attack (CRACK) is explored, where the non-reciprocity in the BD-RIS is utilized to inherently break the reciprocity of uplink (UL) and downlink (DL) channels. Consequently, the performance of DL transmission deteriorates in time-division duplex (TDD) systems that rely on channel reciprocity.

The design of FD wireless communication system is a promising research direction due to its potential to double the communication rates. In FD systems, transceivers can transmit and receive simultaneously on the same frequency band, thereby addressing the challenge of frequency spectrum scarcity confronted by modern communication systems and enhancing the communication rates \cite{sabharwal_band_2014}. Notably, RISs inherently support FD transmission with aligned DL and UL users due to that the aligned DL and UL transmissions are reversible and RISs introduce no noise corruption \cite{basar2019wireless}. These properties have motivated research on RIS-assisted in-band FD systems. In FD case, the primary difficulty lies in the signal of interest (SoI) drowning in the self interference (SI) due to the coupling between transmitter and receiver. As a result, substantial research has focused on boosting the SoI, and thus reducing the effect of SI via RISs \cite{sharma_intelligent_2021, tota_khel_performance_2022}. However, one challenge of the reciprocal BD-RIS and D-RIS is that only \textit{aligned} DL and UL users can be optimally and simultaneously served in FD communication. This is constrained by reciprocal architecture at RIS, such that the impinging and reflected beams can each only probe at one respective direction.

Recently, \cite{li_non-reciprocal_2024} has theoretically analyzed in which condition both DL and UL users can simultaneously achieve the best received power of the signal of interest in a specific single-antenna FD system aided by non-reciprocal BD-RIS. Results are very promising and demonstrated the potential to serve an UL and DL user simultaneously at different directions thanks to non-reciprocal BD-RIS, which would not be possible with a reciprocal BD-RIS (and hence D-RIS). Given the flexibility and limited study of non-reciprocal BD-RIS in FD systems, {we extend those results and focus on sum-rate maximization for non-reciprocal BD-RIS in a general FD scenario considering both direct and reflected links with multi-antennas, multiple DL, and UL users.} {Our focus in this work is on exploring the capability of non-reciprocal BD-RIS in breaking the uplink-downlink channel reciprocity and demonstrating one unique and beneficial application of non-reciprocal BD-RIS. }

\bpara{Contributions and Overview of Results.} 
In this paper, we investigate non-reciprocal BD-RIS in FD systems, comparing it to reciprocal BD-RIS and D-RIS. Our contributions are summarized as follows:

 \begin{enumerate}[leftmargin = *,label =$\bullet$]
    \item We propose a non-reciprocal BD-RIS assisted FD system design. For the case of single DL and UL users, we focus on designing the scattering matrix, which demonstrates the superiority of non-reciprocal BD-RIS in FD systems in terms of DL and UL sum-rates, compared to reciprocal BD-RIS and conventional D-RIS. For the case of multiple DL and UL users, we design not only the scattering matrix, but also the precoder and combiner in the BS to further improve the sum-rates performances. Numerical simulations demonstrate surprising benefits of non-reciprocal BD-RIS:
    \begin{enumerate}[leftmargin = *, label={\arabic*)}]
    \item Non-reciprocal BD-RIS achieves higher DL and UL sum-rates when DL and UL users occupy different locations in both single and multiple DL and UL users cases.  We attribute this benefit to the greater flexibility of non-reciprocal BD-RIS, allowing the impinging and reflected beams to probe in desired directions.
    \item Non-reciprocal BD-RIS achieves greater gains when considering both DL and UL transmissions, but shows smaller gains when the system design prioritizes one-sided transmission.
    \item {Non-reciprocal BD-RIS achieves gains in the presence of structural scattering, which refers to the specular reflection generated by the RIS when it is turned OFF. By explicitly capturing this normally ignored structural scattering, the benefits of non-reciprocal BD-RIS over reciprocal BD-RIS and D-RIS can be better demonstrated.} Specifically, structural scattering degrades sum-rates when DL and UL users are aligned due to strong interference induced in the DL transmission, but it enhances sum-rates when UL users are positioned at conjugate angles relative to the base station.
    % \item The sum-rates increase as the number of RIS elements increases, with non-reciprocal BD-RIS achieving the highest sum-rates compared to the other two types of RISs as the number of RIS elements varies.
    % \item The sum-rates increase as the group size of the group-connected architecture increases, with non-reciprocal BD-RIS achieving the highest sum-rates compared to the other types of RISs as the group size varies.
    \item The sum-rate gain of non-reciprocal BD-RIS over reciprocal BD-RIS and D-RIS increases as the number of RIS element and group size increase. We attribute this to the fact that more reconfigurable RIS elements provide higher flexibility, enabling better probing of the beam in the desired directions. {In addition, we examine the impact of SI on the system performance.}
    \end{enumerate}

    \item We formulate the maximization of FD DL and UL sum-rates as a non-convex optimization problem. The constraints relate to the matrix structures characterized by symmetry, asymmetry and diagonality. We propose a general algorithm to design the precoder and combiner at the BS, and scattering matrix of the BD-RIS. Subsequently, block coordinate descent (BCD) framework and penalty dual decomposition (PDD) are employed to update each block until convergence. This optimization framework considers the maximization of both DL and UL sum-rates and provides a general solution for designing the asymmetric and symmetric matrices, making it different from other BD-RIS optimization methods.

\end{enumerate}

\bpara{Organization of This Paper.} The paper organization is as follows. The system model is described in Section \ref{sec:sys}. We briefly analyze the effect of reciprocity on the received power in Section \ref{sec:reci}. In Section \ref{sec:pro}, the details of sum-rate maximization problem formulation and transformation are introduced. The proposed solution algorithm, and computational complexity are provided in Section \ref{sec:algo}. Section \ref{sec:simu} provides numerical evaluations, and we conclude this work in Section \ref{sec:con}.

\bpara{Notation.} The set of binary elements $\{0, 1\}$, integers, real numbers, and complex numbers are respectively represented by $\mathbb{B}$, $\mathbb{Z}$, $\mathbb{R}$,  and $\mathbb{C}$. Matrices, vectors and scalars are expressed by capital boldface, small boldface and normal fonts, respectively. $\Re(\cdot)$ represent the real part of a complex number.
We denote conjugate, transpose, conjugate-transpose, and inversion of matrix $\mathbf{X}$ by $\mathbf{X}^*$, $\mathbf{X}^\top$,  $\mathbf{X}^H$, and $\mathbf{X}^{-1}$, respectively. The element in the $i^{\mathrm{th}}$ row and $j^{\mathrm{th}}$ column in the matrix $\mathbf{X}$ is denoted by $\mathbf{X}(i,j)$. $\mathbf{I}$ and $\mathbf{0}$ represent identity matrix, and all-zero matrix.
$\operatorname{vec}{(\cdot)}$, $\operatorname{diag}{(\cdot)}$ and $\operatorname{blkdiag}{(\cdot)}$ denote the vectorization operation, diagonal matrix, and block matrix. $\Tr(\cdot)$, $\mathbb{E}{(\cdot)}$ and $\otimes$ denote trace operation, statistical expectation, and Kronecker product. 
\vspace{-5pt}
\section{System model}
\label{sec:sys}
\begin{figure}[t]
    \centering
    \includegraphics[width = 0.4\textwidth]{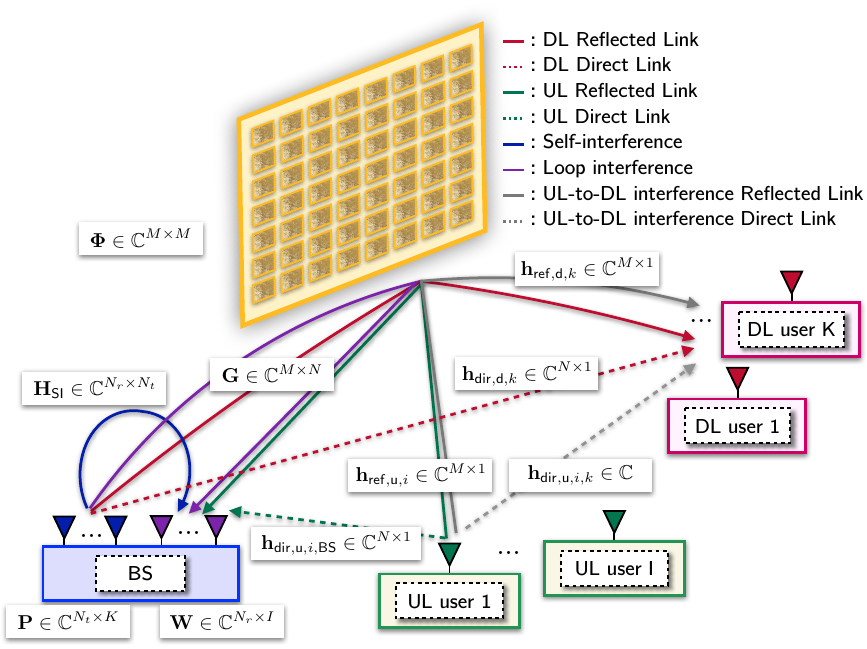}
    \centering
    \caption{{BD-RIS-assisted full-duplex (FD) system model.}}
    \label{fig:system}
    \vspace{-5pt}
\end{figure}
As depicted in \fig{fig:system}, we consider an $M$-element BD-RIS-assisted FD base station (BS) equipped with $N_t$ and $N_r$ transmit and receive antennas, respectively. {We assume that the number of antennas at the BS is $N_t = N_r = N$ for simplicity, and all antenna arrays are uniform linear arrays (ULA) with half-wavelength spacing between adjacent antenna elements. The FD system serves half-duplex (HD) single-antenna DL users and HD UL users simultaneously. The DL and UL users are indexed by $\mathcal{K} = \{1, \dots, K \}$ and $\mathcal{I} = \{1, \dots, I \}$, respectively.} In this system, we assume that perfect instantaneous channel state information (CSI) is available at the BS \cite{li_channel_2024}. We use $\mathbf{\Phi} \in \mathbb{C}^{M \times M}$ to denote the scattering matrix of the BD-RIS. If we consider the group-connected BD-RIS, the group size is denoted by $M_g$, and the total number of groups is $G = M/M_g$. The mathematical expression of the group-connected BD-RIS is expressed as $\mathbf{\Phi} = \operatorname{blkdiag}(\mathbf{\Phi}_1, \cdots, \mathbf{\Phi}_G)$.
% \begin{equation}
%     \mathbf{\Phi} = \operatorname{blkdiag}(\mathbf{\Phi}_1, \cdots, \mathbf{\Phi}_G).
% \end{equation}
Then the single-connected (diagonal) RIS and fully-connected BD-RIS are two extreme case with $G = M$ and $G = 1$.

In the DL direction, the FD BS first precodes the transmit symbol vector $\mathbf{s} \triangleq [s_1, \dots, s_{K}]^\top \in \mathbb{C}^{K}, \mathbb{E}{\{\mathbf{s} \mathbf{s}^H \}} = \mathbf{I}$ using the precoding matrix $\mathbf{P} \triangleq [\mathbf{p}_1, \dots, \mathbf{p}_{K}] \in \mathbb{C}^{N \times K}$, where $\mathbf{p}_{k} \in \mathbb{C}^{N }$ is the precoding vector for the $k^{\mathrm{th}}$ DL user. After pulse shaping and up-conversion, the signal is transmitted through the $N$ antennas and propagates through the direct and reflected channels between the BS and the DL user. Additionally, the DL user receives interference from the UL users through the direct link and the link reflected by the BD-RIS. 

Moreover, we define the effective channels incorporating both direct and reflected links\footnote{Since we utilize the DL and UL channel reciprocity in the FD system, we use transpose $(\cdot)^\top$ instead of Hermitian $(\cdot)^H$ in the channel related expressions to have a more accurate modeling of the channel reciprocity.}. The effective channel between BS and $k^{\mathrm{th}}$ user is defined as
$\mathbf{h}_{\mathsf{d}, k}^\top \triangleq \mathbf{h}_{\mathsf{dir, d}, k}^\top + \mathbf{h}_{\mathsf{ref, d}, k}^\top (\mathbf{\Phi} - \mathbf{I}) \mathbf{G} \in \mathbb{C}^{1 \times N}$. Here, $\mathbf{h}_{\mathsf{dir, d}, k} \in \mathbb{C}^{N \times 1}$ denotes the direct channel between the BS and the $k^{\mathrm{th}}$ DL user. $\mathbf{h}_{\mathsf{ref, d}, k} \in \mathbb{C}^{M \times 1}$ denotes the channel between the BD-RIS and the $k^{\mathrm{th}}$ DL user. 
The channel between BS and the BD-RIS is denoted by $\mathbf{G} \in \mathbb{C}^{M \times N}$. The effective channel between the $i^{\mathrm{th}}$ UL user and the $k^{\mathrm{th}}$ DL user is defined as $\mathbf{h}_{\mathsf{u}, i, k} \triangleq \mathbf{h}_{\mathsf{dir, u}, i, k}^\top + \mathbf{h}_{\mathsf{ref, d}, k}^\top (\mathbf{\Phi} - \mathbf{I}) \mathbf{h}_{\mathsf{ref, u}, i} \in \mathbb{C}$. Here, $\mathbf{h}_{\mathsf{dir, u},i , k} \in \mathbb{C}$ denotes the direct channels between the $i^{\mathrm{th}}$ UL user and the $k^{\mathrm{th}}$ DL user. $\mathbf{h}_{\mathsf{ref, u}, i} \in \mathbb{C}^{M \times 1}$ denotes the channel between the $i^{\mathrm{th}}$ UL user and the BD-RIS. 
Note that we consider the ideal RIS model \cite{10666709} and the terms $-\mathbf{h}_{\mathsf{d},k}^\top\mathbf{G}$ and $-\mathbf{h}_{\mathsf{d},k}^\top \mathbf{h}_{\mathsf{u},i}$ are used to model \textit{structural scattering}. It is usually ignored in most RIS literature, while it has physical meaning that the RIS still radiates even when RIS is turned off due to the non-zero current in RIS elements \cite{10666709, hansen1989relationships, abrardo2024design, li_non-reciprocal_2024, nerini2024universal}. This will cause non-negligible impact on system performance. 
Finally, the received signal at the $k^{\mathrm{th}}$ DL user is:
% \begin{equation}
% \begin{aligned}
% y_{k} 
% % = & \mathbf{h}_{\mathsf{d}, k}^H (\mathbf{\Phi} - \mathbf{I}) \mathbf{G} \mathbf{P} \mathbf{s}+ \sqrt{P_u} \sum_{i \in \mathcal{I}} \mathbf{h}_{\mathsf{d}, k}^H (\mathbf{\Phi} - \mathbf{I}) \mathbf{h}_{\mathsf{u}, i} x_i + n_k, \\
% = & \mathbf{h}_{\mathsf{d}, k}^\top (\mathbf{\Phi} - \mathbf{I}) \mathbf{G p}_k s_k+\mathbf{h}_{\mathsf{d}, k}^\top (\mathbf{\Phi} - \mathbf{I}) \mathbf{G} \sum_{j \in \mathcal{K}, j \neq k} \mathbf{p}_j s_j \\
% & + \underbrace{ \sqrt{P_u} \sum_{i \in \mathcal{I}} \mathbf{h}_{\mathsf{d}, k}^\top (\mathbf{\Phi} - \mathbf{I}) \mathbf{h}_{\mathsf{u}, i} x_i}_{\text{Interference from UL Users}}
% +n_k, \quad  \quad \forall k \in \mathcal{K},
% \label{eq:dlsignal}
% \end{aligned}
% \end{equation}
\begin{equation}
    \begin{aligned}
    y_{k} = & \mathbf{h}_{\mathsf{d}, k}^\top \mathbf{p}_k s_k+\mathbf{h}_{\mathsf{d}, k}^\top \sum_{j \in \mathcal{K}, j \neq k} \mathbf{p}_j s_j \\ +
    & \underbrace{ \sqrt{P_u} \sum_{i \in \mathcal{I}} \mathbf{h}_{\mathsf{u}, i, k} x_i}_{\text{Interference from UL Users}}  +n_k, \quad \forall k \in \mathcal{K},
    \end{aligned}
    \label{eq:dlsignal}
\end{equation}

% where $\mathbf{h}_{\mathsf{d}, k} \in \mathbb{C}^{M \times 1}$, $\mathbf{h}_{\mathsf{u}, i} \in \mathbb{C}^{M \times 1}$, $\mathbf{H}_\mathsf{SI} \in  \mathbb{C}^{N \times N}$, and $\mathbf{G} \in \mathbb{C}^{M \times N}$ denote the channel between the BD-RIS and $k^{\mathrm{th}}$ DL user, the channel between the BD-RIS and $i^{\mathrm{th}}$ UL user, the self-interference (SI) channel, and the channel between the BS and BD-RIS, respectively. 
where $x_i$ represents the transmitted streams from the UL users to the BS. 
% $\mathbf{h}_{\mathsf{d}, k} \in \mathbb{C}^{M}$, $\mathbf{h}_{\mathsf{u}, i} \in \mathbb{C}^{M}$, and $\mathbf{G} \in \mathbb{C}^{M \times N}$ denote the channel between the BD-RIS and $k^{\mathrm{th}}$ DL user, the channel between the BD-RIS and $i^{\mathrm{th}}$ UL user, and the channel between the BS and BD-RIS, respectively. 
We assume the transmit power of each UL user is the same and is denoted by $P_u$. The term $n_k \sim \mathcal{CN}(0, \sigma^2)$ represents the additive Gaussian white noise (AWGN).
% Note that the terms $-\mathbf{h}_{\mathsf{d},k}^\top\mathbf{G}$ and $-\mathbf{h}_{\mathsf{d},k}^\top \mathbf{h}_{\mathsf{u},i}$ are used to model structural scattering. It is usually ignored in most RIS literature, while it has its own physical meaning and will cause non-negligible impact on system performance \cite{hansen1989relationships, abrardo2024design, li_non-reciprocal_2024, nerini2024physics}.

In the UL direction, the FD BS receives the signal from the $i^{\mathrm{th}}$ UL user from both direct link and reflected link by the BD-RIS. In addition, the FD BS receives self-interference (SI) due to the coupling between transmitter and receiver \cite{sabharwal_band_2014}, and the loop interference due to the reflection by the BD-RIS of the transmitted signal. We define the effective channel including direct and reflected links between $i^{\mathrm{th}}$ UL user and BS as 
$\mathbf{h}_{\mathsf{u},i, \mathsf{BS}} \triangleq \mathbf{h}_{\mathsf{dir, u}, i, \mathsf{BS}} + \mathbf{G}^\top (\mathbf{\Phi} - \mathbf{I}) \mathbf{h}_{\mathsf{ref, u}, i} \in \mathbb{C}^{N \times 1}$. Here, $\mathbf{h}_{\mathsf{dir, u}, i, \mathsf{BS}} \in \mathbb{C}^{N \times 1}$ denotes the direct channel between the $i^{\mathrm{th}}$ UL user and the BS.
Finally, the signals are combined using the combiner matrix $\mathbf{W} \triangleq [\mathbf{w}_1, \dots, \mathbf{w}_{I}] \in \mathbb{C}^{N \times I}$. The received signal at the FD BS from the $i^{\mathrm{th}}$ UL is expressed as
% \begin{equation}
% \begin{aligned}
%      z_{i} &
%     % & =  \sqrt{P_u} \mathbf{w}_i^H \mathbf{G}^\top (\mathbf{\Phi} - \mathbf{I}) \mathbf{H}_{\mathsf{u}} \, \mathbf{x} +  \mathbf{w}_i^H \mathbf{H}_\mathsf{SI} \mathbf{P} \mathbf{s} \\
%     % & + \mathbf{w}_i^H \mathbf{G}^\top (\mathbf{\Phi} - \mathbf{I}) \mathbf{G} \mathbf{P} \mathbf{s} + \mathbf{w}_i^H \mathbf{n}_u, \\ 
%     \!=\! \! \sqrt{P_u} \! \mathbf{w}_i^H \mathbf{G}^\top \!(\mathbf{\Phi}\! -\! \mathbf{I}) \mathbf{h}_{\mathsf{u},i} x_i \! +  \! \! \sqrt{P_u} \! \! \! \! \!\sum_{p \in \mathcal{I}, p \neq i } \! \! \! \! \! \! \mathbf{w}_i^H \mathbf{G}^\top \!( \mathbf{\Phi}\! - \!\mathbf{I}) \mathbf{h}_{\mathsf{u},p} x_p \\
%     &+ \underbrace{ \mathbf{w}_i^H \mathbf{H}_\mathsf{SI} \mathbf{P} \mathbf{s}}_{\text{Self-interference}} + \underbrace{\mathbf{w}_i^H \mathbf{G}^\top (\mathbf{\Phi} - \mathbf{I}) \mathbf{G} \mathbf{P} \mathbf{s}}_{\text{Loop interference}} + \mathbf{w}_i^H \mathbf{n}_u, \quad \forall i \in \mathcal{I},
% \end{aligned}
% \label{eq:rxsignal}
% \end{equation}
\begin{equation}
    \begin{aligned}
         z_{i} &
         = \sqrt{P_u} \mathbf{w}_i^H \mathbf{h}_{\mathsf{u},i, \mathsf{BS}} x_i \! +  \! \sqrt{P_u} \sum_{p \in \mathcal{I}, p \neq i }  \mathbf{w}_i^H \mathbf{h}_{\mathsf{u},p, \mathsf{BS}} x_p \\
        &+ \underbrace{ \mathbf{w}_i^H \mathbf{H}_\mathsf{SI} \mathbf{P} \mathbf{s}}_{\text{Self-interference}} + \underbrace{\mathbf{w}_i^H \mathbf{G}^\top (\mathbf{\Phi} - \mathbf{I}) \mathbf{G} \mathbf{P} \mathbf{s}}_{\text{Loop interference}} + \mathbf{w}_i^H \mathbf{n}_u, \quad \forall i \in \mathcal{I},
    \end{aligned}
    \label{eq:rxsignal}
\end{equation}
where $\mathbf{H}_\mathsf{SI} \in  \mathbb{C}^{N \times N}$ denotes the SI channel, and $\mathbf{n}_u \sim \mathcal{CN}(\mathbf{0}, \sigma^2 \mathbf{I})$ represents the AWGN at the FD BS.

% Define $\tilde{\mathbf{h}}_{\mathsf{d}, k} \triangleq (\mathbf{h}_{\mathsf{d}, k}^\top (\mathbf{\Phi-\mathbf{I}} ) \mathbf{G})^\top \in \mathbb{C}^{N \times 1}$, $\tilde{\mathbf{h}}_{\mathsf{u},i} \triangleq  \mathbf{G}^\top (\mathbf{\Phi-\mathbf{I}}) \mathbf{h}_{\mathsf{u},i}  \in \mathbb{C}^{N \times 1}$, 

With the above signal model, the signal-to-interference-plus-noise ratio (SINR) for the $k^{\mathrm{th}}$ DL user and SINR at the FD BS from the $i^{\mathrm{th}}$ UL user is calculated as 
% \begin{equation}
% \gamma_{\mathsf{d}, k}=\frac{\left|\tilde{\mathbf{h}}_{\mathsf{d}, k}^\top \mathbf{p}_k\right|^2}{
% I_{\mathsf{d}}(\mathbf{P}, \mathbf{\Phi})+\sigma^2},
% \label{eq:sinr_dl}
% \end{equation}
\begin{equation}
    \gamma_{\mathsf{d}, k}=\frac{\left|\mathbf{h}_{\mathsf{d}, k}^\top \mathbf{p}_k\right|^2}{
    I_{\mathsf{d}}(\mathbf{P}, \mathbf{\Phi})+\sigma^2},\, \gamma_{\mathsf{u}, i} =\frac{P_u \left|\mathbf{w}_i^H \mathbf{h}_{\mathsf{u}, i, \mathsf{BS}} \right|^2}{ I_{\mathsf{u}}(\mathbf{W}, \mathbf{P}, \mathbf{\Phi}) + \| \mathbf{w}_i \|^2_F \sigma^2}
    \label{eq:sinr_dlandul}
\end{equation}
% \begin{equation}
%     \gamma_{\mathsf{u}, i} =\frac{P_u \left|\mathbf{w}_i^H \mathbf{h}_{\mathsf{u}, i, \mathsf{BS}} \right|^2}{ I_{\mathsf{u}}(\mathbf{W}, \mathbf{P}, \mathbf{\Phi}) + \| \mathbf{w}_i \|^2_F \sigma^2},
% \label{eq:sinr_ul}
% \end{equation}
where the interference power term of the DL and UL transmissions are expressed by
% \begin{equation}
%     I_{\mathsf{d}}(\mathbf{P}, \mathbf{\Phi}) = \sum_{j \in \mathcal{K}, j \neq k}\left|\tilde{\mathbf{h}}_{\mathsf{d}, k}^\top \mathbf{p}_j\right|^2
% + P_u \sum_{i \in \mathcal{I}} \left| \mathbf{h}_{\mathsf{d}, k}^\top (\mathbf{\Phi-\mathbf{I}}) \mathbf{h}_{\mathsf{u}, i} \right|^2
% \end{equation}
\begin{equation}
    I_{\mathsf{d}}(\mathbf{P}, \mathbf{\Phi}) = \sum_{j \in \mathcal{K}, j \neq k}\left|\mathbf{h}_{\mathsf{d}, k}^\top \mathbf{p}_j\right|^2
    + P_u \sum_{i \in \mathcal{I}} \left| \mathbf{h}_{\mathsf{u},i,k} \right|^2
\end{equation}
% \vspace{-10pt}
% \begin{equation}
% \begin{aligned}
% I_{\mathsf{u}}&(\mathbf{W}, \mathbf{P}, \mathbf{\Phi}) =  P_u \sum_{p \in \mathcal{I}, p \neq i}\left|\mathbf{w}_i^H \tilde{\mathbf{h}}_{\mathsf{u}, p} \right|^2\\
% &+ \sum_{k \in \mathcal{K}} | \mathbf{w}_i^H \mathbf{H}_\mathsf{SI} \mathbf{p}_k + \mathbf{w}_i^H \mathbf{G}^\top (\mathbf{\Phi-\mathbf{I}} ) \mathbf{G} \mathbf{p}_k  |^2.
% \end{aligned}
% \end{equation}
\begin{equation}
    \begin{aligned}
    I_{\mathsf{u}}&(\mathbf{W}, \mathbf{P}, \mathbf{\Phi}) =  P_u \sum_{p \in \mathcal{I}, p \neq i}\left|\mathbf{w}_i^H \mathbf{h}_{\mathsf{u}, p, \mathsf{BS}} \right|^2\\
    &+ \sum_{k \in \mathcal{K}} | \mathbf{w}_i^H \mathbf{H}_\mathsf{SI} \mathbf{p}_k + \mathbf{w}_i^H \mathbf{G}^\top (\mathbf{\Phi-\mathbf{I}} ) \mathbf{G} \mathbf{p}_k  |^2.
    \end{aligned}
    \end{equation}

% and $\mathbf{\tilde{\mathbf{H}}}_{\mathsf{u}, p} \triangleq [\mathbf{\tilde{\mathbf{h}}}_{\mathsf{u}, 1}, \cdots, \mathbf{\tilde{\mathbf{h}}}_{\mathsf{u}, I}] \in \mathbb{C}^{N \times I}$.
\vspace{-10pt}
\section{Analysis of the RIS Reciprocity}
% This section is used to explain the observations in the numerical simulations. 
% \subsection{Effect of Reciprocity}
\label{sec:reci}
% As shown in the SINRs expression for the DL user \eqref{eq:sinr_dl} and UL user \eqref{eq:sinr_ul}, the SINRs are mainly affected by the received power term in the numerator and the interference terms in the denominator. 

To show how the RIS reciprocity affects the system performance, we consider a simple scenario where the RIS-assisted FD system serves one DL user and one UL user and all the direct links are blocked. Inspired by \cite{shen_modeling_2022}, we focus on simultaneously maximizing the received power of the signal of interest at both the DL user and the BS.

% Thus, the following problem needs to be solved:
% \begin{maxi!}|s|[2]                   % mini! = minimize 
%     {\mathbf{\Phi}}                               % optimization variable
%     {\eta_\mathsf{d} + \eta_\mathsf{u}. \label{eq:op11}}   % objective function and label
%     {\label{eq:p11}}             % label for opt problem
%     {\mathcal{P}11:} 
%     % optimization result
%     {}
% \end{maxi!}

Starting from maximizing the UL received power, the UL received power of the signal of interest is determined by the channel strength, which is given by
\begin{equation}
    \eta_\mathsf{u} = |\mathbf{g}^\top (\mathbf{\Phi}-\mathbf{I})  \mathbf{h}_{\mathsf{u}, i} |^2.
\end{equation}
 We apply the triangle inequality, which leads to
\begin{equation}
    | \mathbf{g}^\top (\mathbf{\Phi} - \mathbf{I}) \mathbf{h}_{\mathsf{u},i} | \leq | \mathbf{g}^\top \mathbf{\Phi} \mathbf{h}_{\mathsf{u},i} | + | \mathbf{g}^\top \mathbf{h}_{\mathsf{u},i} |.
\label{eq:tri}
\end{equation}
Subsequently, we apply the Cauchy-Schwarz inequality and the unitary condition $\mathbf{\Phi}^H \mathbf{\Phi} = \mathbf{I}$ \cite{shen_modeling_2022} to scale \eqref{eq:tri}. Therefore, the upper bound of $\eta_\mathsf{u}$ is given by
\begin{equation}
    | \mathbf{g}^\top (\mathbf{\Phi} - \mathbf{I}) \mathbf{h}_{\mathsf{u},i} |^2 \leq 
    ( \| \mathbf{g}  \|_2 \| \mathbf{h}_{\mathsf{u},i} \|_2 + | \mathbf{g}^\top \mathbf{h}_{\mathsf{u},i} | )^2.
\label{eq:ul_bound}
\end{equation}
The equality holds true when
\begin{equation}
    \beta_\mathsf{u} \frac{\mathbf{g}^* }{\| \mathbf{g}  \|_2} = \mathbf{\Phi} \frac{\mathbf{h}_{\mathsf{u},i}}{\| \mathbf{h}_{\mathsf{u},i} \|_2}, \, \text{and} \, \, \beta_\mathsf{u} = e^{\jmath \angle\left(-\mathbf{g}^{\top} \mathbf{h}_{\mathsf{u},i}\right)}.
    \label{eq:ul_eq}
\end{equation}
% where $\beta_\mathsf{u} = e^{\jmath \angle\left(-\mathbf{g}^{\top} \mathbf{h}_{\mathsf{u},i}\right)}$. 

In the DL side, the DL received power is also determined by channel strength, which is given by
\begin{equation}
    \eta_\mathsf{d} = |\mathbf{h}_{\mathsf{d}, k}^\top (\mathbf{\Phi}-\mathbf{I})\mathbf{g}|^2.
\end{equation}
We subsequently apply the triangular and Cauchy-Schwarz inequalities to find the upper bound of $\eta_\mathsf{d}$, expressed as
\begin{equation}
    |\mathbf{h}_{\mathsf{d}, k}^\top (\mathbf{\Phi}-\mathbf{I})\mathbf{g}|^2 \leq
    ( \| \mathbf{h}_{\mathsf{d}, k} \|_2 \| \mathbf{g} \|_2  + | \mathbf{h}_{\mathsf{d}, k}^\top \mathbf{g} | )^2,
\end{equation}
where the equality holds true if and only if 
\begin{equation}
    \beta_\mathsf{d} \frac{\mathbf{h}_{\mathsf{d}, k}^* }{\| \mathbf{h}_{\mathsf{d}, k}  \|_2} = \mathbf{\Phi} \frac{\mathbf{g}}{\| \mathbf{g} \|_2}, \, \text{and} \, \, \beta_\mathsf{d} = e^{\jmath \angle\left(-\mathbf{h}_{\mathsf{d},k}^\top \mathbf{g} \right)}.
    \label{eq:dl_eq}
\end{equation}
% and $\beta_\mathsf{d} = e^{\jmath \angle\left(-\mathbf{h}_{\mathsf{d},k}^\top \mathbf{g} \right)}$.

To simultaneously maximize both the DL and UL received power, the optimal solution for $\mathbf{\Phi}$ should satisfy both \eqref{eq:ul_eq} and \eqref{eq:dl_eq}.  To investigate reciprocity in the BD-RIS, we rewrite $ \eta_\mathsf{u}$ by taking the transpose, such that 
$|\mathbf{h}_{\mathsf{u}, i}^\top (\mathbf{\Phi}^\top-\mathbf{I}) \mathbf{g} |^2$, resulting in the equality condition
\begin{equation}
    \beta_{\mathsf{u}} \frac{\mathbf{h}_{\mathsf{u},i}^*}{\| \mathbf{h}_{\mathsf{u},i} \|_2}  = \mathbf{\Phi}^\top \frac{\mathbf{g} }{\| \mathbf{g}  \|_2}.
    \label{eq:ul_eq_t}
\end{equation}
Since $\mathbf{\Phi} = \mathbf{\Phi}^\top$ holds for reciprocal BD-RIS, the equalities \eqref{eq:dl_eq} and \eqref{eq:ul_eq_t} for maximizing  both DL and UL power can be satisfied when 
$\beta_\mathsf{d} {\mathbf{h}_{\mathsf{d}, k}^* }{\| \mathbf{h}_{\mathsf{d}, k}  \|_2}^{-1} = \beta_{\mathsf{u}} {\mathbf{h}_{\mathsf{u},i}^*}{\| \mathbf{h}_{\mathsf{u},i} \|_2}^{-1}$.
% To investigate reciprocity in the BD-RIS, we rewrite $ \eta_\mathsf{d}$ by taking the transpose, such that 
% $|\mathbf{g}^\top (\mathbf{\Phi}^\top-\mathbf{I}) \mathbf{h}_{\mathsf{d}, k}|^2$, resulting in the equality condition
% \begin{equation}
%     \beta_{\mathsf{d}} \frac{\mathbf{g}^* }{\| \mathbf{g}  \|_2} = \mathbf{\Phi}^\top \frac{\mathbf{h}_{\mathsf{d},k}}{\| \mathbf{h}_{\mathsf{d},k} \|_2}.
%     \label{eq:dl_eq_t}
% \end{equation}
% For reciprocal BD-RIS, since $\mathbf{\Phi} = \mathbf{\Phi}^\top$, the equalities \eqref{eq:ul_eq} and \eqref{eq:dl_eq_t} for maximizing  both DL and UL power can be satisfied when $\mathbf{h}_{\mathsf{d}, k}/\| \mathbf{h}_{\mathsf{d},k} \|_2  = \mathbf{h}_{\mathsf{u}, i}/\| \mathbf{h}_{\mathsf{u},i} \|_2$, and $\beta_\mathsf{d} = \beta_\mathsf{u}$.
This condition requires the normalized DL and UL channels to be co-linear, which only occurs when the DL and UL users are aligned and experience identical small-scale fading. Consequently, reciprocal BD-RIS cannot support simultaneous DL and UL transmission for non-aligned DL and UL users. 
In contrast, non-reciprocal BD-RIS, which does not impose a symmetry constraint (\ie $\mathbf{\Phi} \neq \mathbf{\Phi}^\top$), provides greater flexibility in the solution space to maximize both DL and UL received power. This theoretical advantage has been demonstrated in \cite{li_non-reciprocal_2024}. Building on this insight, the following sections explore a more complex and general scenario involving multiple DL and UL users. We formulate a sum-rate maximization problem, propose a solution, and verify whether the benefits of non-reciprocal BD-RIS persist in these settings.

\section{Problem Formulation and Transformation}
\label{sec:pro}
{In this section, we formulate the optimization problem for maximizing the DL and UL sum-rates in the FD system with both direct and reflected links.} We then address the non-convexity of the formulated problem. Due to the intractability of the fractional structure within the $\log(\cdot)$ function, we apply the Lagrangian dual transform to separate this term from the $\log(\cdot)$ function. Next, we employ the quadratic transform to further transform the fractional components into integral expressions \cite{shen2018fractional1, shen2018fractional2}. The details of the problem formulation and transformation are provided below.

\subsection{Problem Formulation}
Our objective is to maximize the DL and UL sum rates in this RIS-aided FD system. To achieve this, we formulate a sum-rate maximization problem and optimize the precoder matrix $\mathbf{P}$, the receive combiner matrix $\mathbf{W}$, and the RIS scattering matrix $\mathbf{\Phi}$. The weighted DL and UL sum-rates are:
\begin{equation}
    f_o(\mathbf{P}, \mathbf{W}, \mathbf{\Phi}) \! \triangleq \! \alpha_\mathsf{d} \! \sum_{k \in \mathcal{K}} \log_2 (1+\gamma_{\mathsf{d}, k})  +  \alpha_\mathsf{u} \! \sum_{i \in \mathcal{I}} \log_2 (1+\gamma_{\mathsf{u}, i}),
\end{equation}
where $\alpha_\mathsf{d}$ and $\alpha_\mathsf{u}$ are the priority of DL and UL communications, and $\alpha_\mathsf{d}+\alpha_\mathsf{u} = 1$. Then, we have
\begin{maxi!}|s|[2]                   % mini! = minimize 
    {\mathbf{P}, \mathbf{W}, \mathbf{\Phi}}                               % optimization variable
    {f_o(\mathbf{P}, \mathbf{W}, \mathbf{\Phi}) \label{eq:op1}}   % objective function and label
    {\label{eq:p1}}             % label for opt problem
    {\mathcal{P}1:} 
    % optimization result
    {} 
    \addConstraint{\|\mathbf{P}\|^2_F}{\leq P_d, \label{eq:op1c1}}
    \addConstraint{\|\mathbf{W}\|^2_F}{= 1, \label{eq:op1c2}}
    \addConstraint{\mathbf{\Phi}}{\in\mathcal{R}_i, \, i \in\{1,2\}, \label{eq:op1c4}}
    \addConstraint{\mathbf{\Phi}}{\in \mathcal{S}_\ell, \, \ell \in\{1,2,3\}. \label{eq:op1c3}}        
\end{maxi!}
Constraint \eqref{eq:op1c1} and \eqref{eq:op1c2} limit the total transmit and receive power of the FD BS, respectively, with $P_d$ representing the power budget at the transmit side. Constraint \eqref{eq:op1c4} guarantees the scattering matrix is symmetric for reciprocal BD-RIS, \ie $\mathbf{\Phi}\in\mathcal{R}_1 = \{\mathbf{\Phi}  |  \mathbf{\Phi} = \mathbf{\Phi}^\mathsf{T}\}$ and asymmetric for non-reciprocal BD-RIS, \ie $\mathbf{\Phi}\in\mathcal{R}_2 = \{\mathbf{\Phi}  | \mathbf{\Phi} \ne \mathbf{\Phi}^\mathsf{T}\}$. Constraint \eqref{eq:op1c3} ensures losslessness in the multi-port network \cite{shen_modeling_2022, pozar_microwave_2021}, such that $\mathbf{\Phi}$ is constrained by: \textit{i}) $\mathcal{S}_1 = \{ \mathbf{\Phi} = \text{diag}(\phi_1, \ldots,\phi_M) | |\phi_m | = 1, \quad \forall m \in \mathcal{M} \}$ for the single-connected (diagonal) RIS, \textit{ii}) $\mathcal{S}_2 = \{\mathbf{\Phi} = \operatorname{blkdiag}(\mathbf{\Phi}_1, \cdots, \mathbf{\Phi}_G) | \boldsymbol{\Phi}_{g}^H \boldsymbol{\Phi}_{ g}= \mathbf{I}, \quad \forall g \in \mathcal{G} \}$ for the group-connected BD-RIS, and \textit{iii}) $\mathcal{S}_3 = \{\mathbf{\Phi} |  \mathbf{\Phi}^H \mathbf{\Phi} = \mathbf{I} \}$ for the fully-connected BD-RIS. Problem $\mathcal{P}1$ includes an intractable $\log(\cdot)$ term and a fractional structure in the objective function, along with non-convex constraints. These conditions result in difficulty in a direct solution. To address this difficulty, we use a fractional programming-based method \cite{shen2018fractional1, shen2018fractional2} to make $\mathcal{P}1$ more tractable and solve it using an iterative method.

\subsection{Problem Transformation}
\bpara{Lagrangian Dual Transformation.} Since the fractional term is challenging to tackle with, we separate it out of the $\log(\cdot)$  from the objective function \eqref{eq:op1} by Lagrangian dual transformation \cite{shen2018fractional1, shen2018fractional2}. This introduces a summation over a new fractional term. The transformed function is given by
\begin{equation}
    \begin{aligned}
        &f_\iota(\mathbf{P},\! \mathbf{W},\! \mathbf{\Phi},\! \boldsymbol{\iota}) \! \! = \\
        & \quad \alpha_\mathsf{d}  \sum_{k \in \mathcal{K}} \! \! \Bigg( \!  \! \log_2(1 \! + \!\iota_{\mathsf{d}, k}) \! - \! \iota_{\mathsf{d}, k} \! + \! \frac{(1+\iota_{\mathsf{d}, k}) \left|{\mathbf{h}}_{\mathsf{d}, k}^\top \mathbf{p}_k\right|^2}{\Gamma_{\mathsf{d}} \!+ \!\sigma^2}  \! \Bigg) \\
        & + \alpha_\mathsf{u} \! \sum_{i \in \mathcal{I}} \! \! \Bigg( \!\!  \log_2(1+\iota_{\mathsf{u}, i}) \! - \! \iota_{\mathsf{u}, i} + \frac{(1+\iota_{\mathsf{u}, i}) P_u \left|\mathbf{w}_i^H {\mathbf{h}}_{\mathsf{u}, i, \mathsf{BS}} \right|^2}{\Gamma_{\mathsf{u}} \! + \! \| \mathbf{w}_i \|^2_F \sigma^2} \!  \Bigg) \!,
    \end{aligned}
\end{equation}
% \begin{figure*}
% \begin{equation}
%     \begin{aligned}
%         f_\iota(\mathbf{P},\! \mathbf{W},\! \mathbf{\Phi},\! \boldsymbol{\iota}) \! \! = 
%         \alpha_\mathsf{d}  \sum_{k \in \mathcal{K}} \! \! \Bigg( \!  \! \log_2(1 \! + \!\iota_{\mathsf{d}, k}) \! - \! \iota_{\mathsf{d}, k} \! + \! \frac{(1+\iota_{\mathsf{d}, k}) \left|{\mathbf{h}}_{\mathsf{d}, k}^\top \mathbf{p}_k\right|^2}{\Gamma_{\mathsf{d}} \!+ \!\sigma^2}  \! \Bigg) 
%         + \alpha_\mathsf{u} \! \sum_{i \in \mathcal{I}} \! \! \Bigg( \!\!  \log_2(1+\iota_{\mathsf{u}, i}) \! - \! \iota_{\mathsf{u}, i} + \frac{(1+\iota_{\mathsf{u}, i}) P_u \left|\mathbf{w}_i^H {\mathbf{h}}_{\mathsf{u}, i, \mathsf{BS}} \right|^2}{\Gamma_{\mathsf{u}} \! + \! \| \mathbf{w}_i \|^2_F \sigma^2} \!  \Bigg) \!,
%     \end{aligned}
% \end{equation}
% \end{figure*}
where $\boldsymbol{\iota} \triangleq [\boldsymbol{\iota}_{\mathsf{d}}^\top, \boldsymbol{\iota}_{\mathsf{u}}^\top]^\top = [\iota_{\mathsf{d}, 1}, \cdots, \iota_{\mathsf{d}, K}, \iota_{\mathsf{u}, 1}, \cdots, \iota_{\mathsf{u}, I}]^\top \in \mathbb{R}^{K+I}$ is the auxiliary vector. 
$\Gamma_{\mathsf{d}}$ and $\Gamma_{\mathsf{u}}$ are defined by
\begin{equation}
    \Gamma_{\mathsf{d}} = I_{\mathsf{d}}(\mathbf{P}, \mathbf{\Phi}) + \left|{\mathbf{h}}_{\mathsf{d}, k}^\top \mathbf{p}_k\right|^2,
\end{equation}
\begin{equation}
    \Gamma_{\mathsf{u}}  = I_{\mathsf{u}}(\mathbf{W}, \mathbf{P}, \mathbf{\Phi}) + 
    P_u \left|\mathbf{w}_i^H {\mathbf{h}}_{\mathsf{u}, i, \mathsf{BS}} \right|^2.
\end{equation}

% \begin{equation}
%     \Gamma_{\mathsf{d}} = {\sum_{j \in \mathcal{K}}}\left|\tilde{\mathbf{h}}_{\mathsf{d}, k}^\top \mathbf{p}_j\right|^2
%     + P_u \sum_{i \in \mathcal{I}} \left| \mathbf{h}_{\mathsf{d}, k}^\top {(\mathbf{\Phi-\mathbf{I}})} \mathbf{h}_{\mathsf{u}, i} \right|^2,
% \end{equation}
% \begin{equation}
% \begin{aligned}
%     \Gamma_{\mathsf{u}}  = & P_u {\sum_{p \in \mathcal{I}}}\left|\mathbf{w}_i^H \tilde{\mathbf{h}}_{\mathsf{u}, p} \right|^2 + \sum_{j \in \mathcal{K}} | \mathbf{w}_i^H \mathbf{H}_\mathsf{SI} \mathbf{p}_k |^2 \\
%     & + \sum_{j \in \mathcal{K}} | \mathbf{w}_i^H \mathbf{G}^\top {(\mathbf{\Phi-\mathbf{I}} )} \mathbf{G} \mathbf{p}_k  |^2.
% \end{aligned}
% \end{equation}
% \vspace{-5pt}
\textbf{Quadratic Transformation.} The fractional term is still not tractable, thus we utilize quadratic transformation \cite{shen2018fractional1, shen2018fractional2} to transform these components to integral expressions. The reformulated objective function is expressed as 
\begin{equation}
\begin{aligned}
    &f_\tau(\mathbf{P}, \mathbf{W}, \mathbf{\Phi},\boldsymbol{\iota}, \boldsymbol{\tau}) =  \\
    & \alpha_\mathsf{d} \! \sum_{k \in \mathcal{K}} \! \! \Big( \log_2(1+\iota_{\mathsf{d}, k})\! -\! \iota_{\mathsf{d}, k} \! + \!{2\sqrt{1+\iota_{\mathsf{d}, k}} \Re{\tau_{\mathsf{d}, k}^* {\mathbf{h}}_{\mathsf{d}, k}^\top \mathbf{p}_k}} \! \\
    & - \! |\tau_{\mathsf{d}, k}|^2 (\Gamma_{\mathsf{d}} + \sigma^2) \! \Big)  + \alpha_\mathsf{u} \sum_{i \in \mathcal{I}} \Big(  \log_2(1+\iota_{\mathsf{u}, i}) - \iota_{\mathsf{u}, i} \\
    & + \! 2 \sqrt{(1+\iota_{\mathsf{u}, i}) P_u} \!  \Re{ \tau_{\mathsf{u}, i}^* \mathbf{w}_i^H {\mathbf{h}}_{\mathsf{u}, i, \mathsf{BS}}}  \! \! -  \! |\tau_{\mathsf{u}, i}|^2 (\Gamma_{\mathsf{u}}  \! +  \! \| \mathbf{w}_i \|^2_F \sigma^2)  \! \Big)\!,
\end{aligned}
\label{eq:ftau}
\end{equation}
where $\boldsymbol{\tau} \triangleq [\boldsymbol{\tau}_{\mathsf{d}}^\top, \boldsymbol{\tau}_{\mathsf{u}}^\top]^\top = [\tau_{\mathsf{d}, 1}, \cdots, \tau_{\mathsf{d}, K}, \tau_{\mathsf{u}, 1}, \cdots, \tau_{\mathsf{u}, I}]^\top \in \mathbb{R}^{K+I}$ is another introduced auxiliary vector. 

After the above transformations, the original problem $\mathcal{P}1$ is reformulated to
\begin{maxi!}|s|[2]                   % mini! = minimize 
    {\substack{\mathbf{\Phi}, \mathbf{P}, \mathbf{W},\\ \boldsymbol{\iota}, \boldsymbol{\tau}}}                               % optimization variable
    {f_\tau(\mathbf{P}, \mathbf{W}, \mathbf{\Phi},\boldsymbol{\iota}, \boldsymbol{\tau}) \label{eq:op2}}   % objective function and label
    {\label{eq:p2}}             % label for opt problem
    {\mathcal{P}2:} 
    % optimization result eq:op1c1
    {} 
    \addConstraint{\eqref{eq:op1c1}, \eqref{eq:op1c2}, \eqref{eq:op1c4}, \text{and}\, \eqref{eq:op1c3}}{}
    % \addConstraint{\|\mathbf{P}\|^2_F}{\leq P_d, \label{eq:op2c1}}
    % \addConstraint{\|\mathbf{W}\|^2_F}{= 1, \label{eq:op2c2}}
    % \addConstraint{\mathbf{\Phi}}{\in\mathcal{R}_i, \, i \in\{1,2\}, \, \text{select one value for } i\label{eq:op2c3}}
    % \addConstraint{\mathbf{\Phi}}{\in \mathcal{S}_\ell, \, \ell \in \{ 1,2,3 \},\, \text{select one value for } \ell \label{eq:op2c4}}.
\end{maxi!}

\vspace{-10pt}
\section{Solution to DL and UL Sum-rate Maximization}
\label{sec:algo}
To tackle with the unitary constraint in the optimization problem, we transform the problem based on PDD method. To solve this reformulated multi-variable problem, we adopt the BCD framework \cite{li2022beyond} to iteratively update each variable until the objective function converges. The details of the algorithm are given below. The transformed problem, $\mathcal{P}2$, is a typical multi-variable problem, which can be efficiently solved by the BCD iterative algorithms \cite{bertsekas1997nonlinear}. Specifically, we first initialize the optimization variables (\ie $\mathbf{P}, \mathbf{W}, \mathbf{\Phi}$), then each variable is updated while keeping the others fixed until convergence. The proposed design algorithm is summarized in Algorithm \ref{alg:alg1}. To make the expressions compact, we denote all the individual channels by $\mathbf{H}_{\mathsf{d}} \triangleq [\mathbf{h}_{\mathsf{d}, 1}, \cdots, \mathbf{h}_{\mathsf{d}, K}] \in \mathbb{C}^{M \times K}$, $\mathbf{H}_{\mathsf{u, BS}} \triangleq [\mathbf{h}_{\mathsf{u}, 1, \mathsf{BS}}, \cdots, \mathbf{h}_{\mathsf{u}, I, \mathsf{BS}}] \in \mathbb{C}^{M \times I}$, and $\mathbf{H}_{\mathsf{u, DL}} \triangleq [\mathbf{h}_{\mathsf{u}, 1, 1}, \cdots, \mathbf{h}_{\mathsf{u}, I, K}] \in \mathbb{C}^{(M \times I) \times 1}$ respectively.
\begin{algorithm}[t]
	\caption{Proposed Algorithm for FD DL and UL Sum-rates Design}
	\label{alg:alg1}
	\KwIn{$\mathbf{H}_{\mathsf{d}}, \mathbf{H}_{\mathsf{u, BS}}, \mathbf{H}_{\mathsf{u, DL}}, \mathbf{H}_\mathsf{SI}, \mathbf{G}$.}  
	\KwOut{$\mathbf{\Phi}^\mathsf{opt}, \mathbf{P}^\mathsf{opt}, \mathbf{W}^\mathsf{opt}$.} 
	\BlankLine
	Initialize $\mathbf{\Phi}, \mathbf{P}, \mathbf{W}, t=1$.
	
	\While{\textnormal{no convergence of objective function \eqref{eq:op2} \textbf{\&}} $\quad t<t_\mathsf{max}$ }{
        Update $\boldsymbol{\iota}_\mathsf{d}^\mathsf{opt}$ and $\boldsymbol{\iota}_\mathsf{u}^\mathsf{opt}$ by \eqref{eq:sinr_dlandul}, respectively. \\
        Update $\boldsymbol{\tau}_\mathsf{d}^\mathsf{opt}$ and $\boldsymbol{\tau}_\mathsf{u}^\mathsf{opt}$ by \eqref{eq:taudandu}, respectively. \\
        Update $\mathbf{P}^\mathsf{opt}$ by \eqref{eq:poptimal}. \\
        Update $\mathbf{W}^\mathsf{opt}$ by \eqref{eq:woptimal}. \\
		Update $\mathbf{\Phi}^\mathsf{opt}$ by Algorithm \ref{alg:alg2}. \\
        $t = t + 1$.\\
	}	
	Return $\mathbf{\Phi}^\mathsf{opt}, \mathbf{P}^\mathsf{opt}, \mathbf{W}^\mathsf{opt}$.
\end{algorithm}
In the following subsections, we provide the details of decomposition of $\mathcal{P}1$ to several sub-problems. Additionally, the derivations of the optimal solution of each block are given. 

\bpara{Auxiliary Vectors: Block $\boldsymbol{\iota}$.} 
When $\mathbf{P}, \mathbf{W}, \mathbf{\Phi}, \boldsymbol{\tau}$ are all fixed, the two separate sub-problems with regards to $\boldsymbol{\iota}_{\mathsf{d}}$ and $\boldsymbol{\iota}_{\mathsf{u}}$ are both convex and unconstrained. Thus, we can obtain the optimal by taking the derivatives with respect to (w.r.t) these two variables and setting to $\mathbf{0}$, \ie $\frac{\partial f_\tau(\mathbf{P}, \mathbf{W}, \mathbf{\Phi},\boldsymbol{\iota}, \boldsymbol{\tau})}{\partial \boldsymbol{\iota}_{\mathsf{d}}} = \mathbf{0}$, or $\frac{\partial f_\tau(\mathbf{P}, \mathbf{W}, \mathbf{\Phi},\boldsymbol{\iota}, \boldsymbol{\tau})}{\partial \boldsymbol{\iota}_{\mathsf{u}}} = \mathbf{0}$. Therefore, the optimal $\boldsymbol{\iota}_{\mathsf{d}}^\mathsf{opt}$ and $\boldsymbol{\iota}_{\mathsf{u}}^\mathsf{opt}$ have the same expressions as the DL and UL SINR in \eqref{eq:sinr_dlandul}, respectively

% Therefore, the optimal $\boldsymbol{\iota}_{\mathsf{d}}^\mathsf{opt}$ and $\boldsymbol{\iota}_{\mathsf{u}}^\mathsf{opt}$ are:
% \begin{equation}
%     \iota_{\mathsf{d}, k}^\mathsf{opt} = \gamma_{\mathsf{d}, k}=\frac{\left|{\mathbf{h}}_{\mathsf{d}, k}^\top \mathbf{p}_k\right|^2}{I_{\mathsf{d}}(\mathbf{P}, \mathbf{\Phi})
%     +\sigma^2}, \quad \forall k \in \mathcal{K},
%     \label{eq:iotad}
% \end{equation}
% \begin{equation}
%     \iota_{\mathsf{u},i}^\mathsf{opt} = \gamma_{\mathsf{u}, i} =\frac{P_u \left|\mathbf{w}_i^H {\mathbf{h}}_{\mathsf{u}, i, \mathsf{BS}} \right|^2}{ I_{\mathsf{u}}(\mathbf{W}, \mathbf{P}, \mathbf{\Phi}) + \| \mathbf{w}_i \|^2_F \sigma_k^2}, \quad \forall i \in \mathcal{I}.
%     \label{eq:iotau}
% \end{equation}

\bpara{Auxiliary Vectors: Block $\boldsymbol{\tau}$.} Similar to the update process of blocks $\boldsymbol{\iota}$, the two sub problems are unconstrained convex optimization under the fixed $\mathbf{P}, \mathbf{W}, \mathbf{\Phi}, \boldsymbol{\iota}$. As a result, the optimal $\boldsymbol{\tau}_\mathsf{d}^\mathsf{opt}$ and $\boldsymbol{\tau}_\mathsf{u}^\mathsf{opt}$ are obtained by $\frac{\partial f_\tau(\mathbf{P}, \mathbf{W}, \mathbf{\Phi},\boldsymbol{\iota}, \boldsymbol{\tau})}{\partial \boldsymbol{\tau}_{\mathsf{d}}} = \mathbf{0}$, or $\frac{\partial f_\tau(\mathbf{P}, \mathbf{W}, \mathbf{\Phi},\boldsymbol{\iota}, \boldsymbol{\tau})}{\partial \boldsymbol{\tau}_{\mathsf{u}}} = \mathbf{0}$. The expressions are given by
% \begin{equation}
% \tau_{\mathsf{d}, k}^\mathsf{opt} = \frac{ \sqrt{1+\iota_{\mathsf{d}, k}}{\mathbf{h}}_{\mathsf{d}, k}^\top \mathbf{p}_k}
% {\Gamma_\mathsf{d}
% +\sigma^2},
% \label{eq:taud}
% \end{equation}
% \vspace{-5pt}
% \begin{equation}
% \tau_{\mathsf{u},i}^\mathsf{opt} = \frac{\sqrt{(1+\iota_{\mathsf{u}, i})} \sqrt{P_u} \mathbf{w}_i^H {\mathbf{h}}_{\mathsf{u}, i, \mathsf{BS}} }
% {\Gamma_\mathsf{u} + \| \mathbf{w}_i \|^2_F \sigma^2}.
% \label{eq:tauu}
% \end{equation}
\begin{equation}
\tau_{\mathsf{d}, k}^\mathsf{opt} = \frac{ \sqrt{1+\iota_{\mathsf{d}, k}}{\mathbf{h}}_{\mathsf{d}, k}^\top \mathbf{p}_k}
{\Gamma_\mathsf{d}
+\sigma^2}, \tau_{\mathsf{u},i}^\mathsf{opt} = \frac{\sqrt{(1+\iota_{\mathsf{u}, i})} \sqrt{P_u} \mathbf{w}_i^H {\mathbf{h}}_{\mathsf{u}, i, \mathsf{BS}} }
{\Gamma_\mathsf{u} + \| \mathbf{w}_i \|^2_F \sigma^2}.
\label{eq:taudandu}
\end{equation}
% \vspace{-5pt}
% \begin{equation}
% \tau_{\mathsf{u},i}^\mathsf{opt} = \frac{\sqrt{(1+\iota_{\mathsf{u}, i})} \sqrt{P_u} \mathbf{w}_i^H {\mathbf{h}}_{\mathsf{u}, i, \mathsf{BS}} }
% {\Gamma_\mathsf{u} + \| \mathbf{w}_i \|^2_F \sigma^2}.
% \label{eq:tauu}
% \end{equation}

\bpara{Transmit Precoder: Block $\mathbf{P}$.} Under the condition that $\mathbf{W}, \mathbf{\Phi},\boldsymbol{\iota}$, and $\boldsymbol{\tau}$ are all fixed, we extract the terms related to $\mathbf{P}$, which are expressed by
% \begin{equation}
% \begin{aligned}
%      &f_\tau(\mathbf{P}) \!  =   \alpha_\mathsf{d} \sum_{k \in \mathcal{K}} \Big( \! {2 \sqrt{1+\iota_{\mathsf{d}, k}} \Re{\tau_{\mathsf{d}, k}^* \tilde{\mathbf{h}}_{\mathsf{d}, k}^\top \mathbf{p}_k}} 
%     \! \\
%     & - \! \mathbf{p}_k^H (\sum_{ p \in \mathcal{K}} |\tau_{\mathsf{d}, k}|^2 \tilde{\mathbf{h}}_{\mathsf{d}, p}^* \tilde{\mathbf{h}}_{\mathsf{d}, p}^\top) \mathbf{p}_k \Big) \! - \!  \alpha_\mathsf{u} \Big(\! \sum_{i \in \mathcal{I}} |\tau_{\mathsf{u}, i}|^2 \\
%     & \quad \quad \sum_{k \in \mathcal{K}} |\mathbf{w}_i^H \mathbf{H}_\mathsf{SI} \mathbf{p}_k + \mathbf{w}_i^H \mathbf{G}^\top (\mathbf{\Phi}-\mathbf{I}) \mathbf{G} \mathbf{p}_k |^2  \Big).
%     \end{aligned}
% \end{equation}
\begin{equation}
    \begin{aligned}
         &f_\tau(\mathbf{P}) \!  =   \alpha_\mathsf{d} \sum_{k \in \mathcal{K}} \Big( \! {2 \sqrt{1+\iota_{\mathsf{d}, k}} \Re{\tau_{\mathsf{d}, k}^* {\mathbf{h}}_{\mathsf{d}, k}^\top \mathbf{p}_k}} 
        \! \\
        & - \! \mathbf{p}_k^H (\sum_{ p \in \mathcal{K}} |\tau_{\mathsf{d}, k}|^2 {\mathbf{h}}_{\mathsf{d}, p}^* {\mathbf{h}}_{\mathsf{d}, p}^\top) \mathbf{p}_k \Big) \! - \!  \alpha_\mathsf{u} \Big(\! \sum_{i \in \mathcal{I}} |\tau_{\mathsf{u}, i}|^2 \\
        & \quad \quad \sum_{k \in \mathcal{K}} |\mathbf{w}_i^H \mathbf{H}_\mathsf{SI} \mathbf{p}_k + \mathbf{w}_i^H \mathbf{G}^\top (\mathbf{\Phi}-\mathbf{I}) \mathbf{G} \mathbf{p}_k |^2  \Big).
        \end{aligned}
\end{equation}
The sub-problem w.r.t $\mathbf{P}$ is given by
\begin{maxi!}|s|[2]                   % mini! = minimize 
    {\mathbf{P}}                               % optimization variable
    {f_\tau(\mathbf{P}) \label{eq:op3}}   % objective function and label
    {\label{eq:p3}}             % label for opt problem
    {\mathcal{P}3:} 
    % optimization result
    {} 
    \addConstraint{\|\mathbf{P}\|^2_F}{\leq P_d \label{eq:op3c1}}.
\end{maxi!}

Since the objective function \eqref{eq:op3} and constraint \eqref{eq:op3c1} are all convex, the Lagrange multiplier method based on Karush–Kuhn–Tucker (KKT) condition can be used to obtain the optimum. The expression with an introduced multiplier $\mu$ is given below
\begin{maxi!}|s|[2]                   % mini! = minimize 
    {\mathbf{P}}                               % optimization variable
    {f_\tau(\mathbf{P}) - \mu (\|\mathbf{P}\|^2_F - P_d). \label{eq:op4}}   % objective function and label
    {\label{eq:p4}}             % label for opt problem
    {\mathcal{P}4:} 
    % optimization result
    {} 
\end{maxi!}
Subsequently, we take the partial derivative of the Lagrangian function with regard to $\mathbf{P}$ and $\mu$, respectively, and set both to $\mathbf{0}$. We can obtain the optimal solution of each precoder $\mathbf{p}_k$:
% \begin{equation}
% \begin{aligned}
%     {\mathbf{p}_k^\mathsf{opt}} & \! = \! \Big( \!
%     \alpha_\mathsf{d} \sum_{ p \in \mathcal{K}} \! |\tau_{\mathsf{d}, k}|^2 \tilde{\mathbf{h}}_{\mathsf{d}, p}^* \tilde{\mathbf{h}}_{\mathsf{d}, p}^\top \!+ \! \alpha_\mathsf{u} \Big( \! \sum_{i \in \mathcal{I}} \! |\tau_{\mathsf{u}, i}|^2 \big( (\mathbf{H}_\mathsf{SI}^H \mathbf{w}_i \mathbf{w}_i^H \mathbf{H}_\mathsf{SI}) \\
%     &+ 2 \Re{\mathbf{H}_\mathsf{SI}^H \mathbf{w}_i \mathbf{w}_i^H \mathbf{G}^\top (\mathbf{\Phi} - \mathbf{I})\mathbf{G}}\\
%     & + \mathbf{G}^H (\mathbf{\Phi} - \mathbf{I})^H \mathbf{G}^* \mathbf{w}_i \mathbf{w}_i^H \mathbf{G}^\top (\mathbf{\Phi} - \mathbf{I}) \mathbf{G} \big) \Big) + \mu^\mathsf{opt} \mathbf{I} \Big)^{-1} \\
%     & \alpha_\mathsf{d} \sqrt{1+\iota_{\mathsf{d}, k}} \tau_{\mathsf{d}, k} \tilde{\mathbf{h}}_{\mathsf{d}, k}^*, \quad k \in \mathcal{K},
% \end{aligned}
% \label{eq:poptimal}
% \end{equation}
\begin{equation}
    \begin{aligned}
        {\mathbf{p}_k^\mathsf{opt}} & \! = \! \Big( \!
        \alpha_\mathsf{d} \sum_{ p \in \mathcal{K}} \! |\tau_{\mathsf{d}, k}|^2 {\mathbf{h}}_{\mathsf{d}, p}^* {\mathbf{h}}_{\mathsf{d}, p}^\top \!+ \! \alpha_\mathsf{u} \Big( \! \sum_{i \in \mathcal{I}} \! |\tau_{\mathsf{u}, i}|^2 \big( (\mathbf{H}_\mathsf{SI}^H \mathbf{w}_i \mathbf{w}_i^H \mathbf{H}_\mathsf{SI}) \\
        &+ 2 \Re{\mathbf{H}_\mathsf{SI}^H \mathbf{w}_i \mathbf{w}_i^H \mathbf{G}^\top (\mathbf{\Phi} - \mathbf{I})\mathbf{G}}\\
        & + \mathbf{G}^H (\mathbf{\Phi} - \mathbf{I})^H \mathbf{G}^* \mathbf{w}_i \mathbf{w}_i^H \mathbf{G}^\top (\mathbf{\Phi} - \mathbf{I}) \mathbf{G} \big) \Big) + \mu^\mathsf{opt} \mathbf{I} \Big)^{-1} \\
        & \alpha_\mathsf{d} \sqrt{1+\iota_{\mathsf{d}, k}} \tau_{\mathsf{d}, k} {\mathbf{h}}_{\mathsf{d}, k}^*, \quad k \in \mathcal{K},
    \end{aligned}
    \label{eq:poptimal}
\end{equation}
where $\mu^\mathsf{opt}$ can be obtained through the bisection search.

\bpara{Receive Combiner: Block $\mathbf{W}$.}
With determined $\mathbf{P}, \mathbf{\Phi},\boldsymbol{\iota}$, and $\boldsymbol{\tau}$, we extract the components w.r.t $\mathbf{W}$ in \eqref{eq:ftau}:
\begin{equation}
\begin{aligned}
    f_\tau(\mathbf{W}) &  = \alpha_\mathsf{u} \sum_{i \in \mathcal{I}} \Bigg( 2 \sqrt{1+\iota_{\mathsf{u}, i}} \sqrt{P_u} \Re{\tau_{\mathsf{u}, i}^* \mathbf{w}_i^H {\mathbf{h}}_{\mathsf{u},i, \mathsf{BS}}} \\
    & - | \tau_{\mathsf{u}, i} |^2 \bigg( P_u \sum_{p \in \mathcal{I}} |\mathbf{w}_i^H  {\mathbf{h}}_{\mathsf{u},p, \mathsf{BS}} |^2 + \sum_{k \in \mathcal{K}} | \mathbf{w}_i^H \mathbf{H}_\mathsf{SI} \mathbf{p}_k\\
    & +\mathbf{w}_i^H \mathbf{G}^\top (\mathbf{\Phi} - \mathbf{I}) \mathbf{G} \mathbf{p}_k |^2 + \|\mathbf{w}_i\|^2 \sigma^2 \bigg)
    \Bigg).
\end{aligned}
\vspace{-10pt}
\end{equation}
The sub-problem is thus given by
\begin{maxi!}|s|[2]                   % mini! = minimize 
    {\mathbf{W}}                               % optimization variable
    {f_\tau(\mathbf{W}) \label{eq:op5}}   % objective function and label
    {\label{eq:p5}}             % label for opt problem
    {\mathcal{P}5:} 
    % optimization result
    {}
    \addConstraint{\|\mathbf{W}\|^2_F=1.}{\label{eq:op5c1}}
\end{maxi!}
Although the constraint \eqref{eq:op5c1} is non-convex, we first address this sub-problem as an unconstrained complex optimization problem w.r.t $\mathbf{W}$. Once the algorithm converges, we normalize the obtained solution. The optimal solution of each combiner $\mathbf{w}_i$ is obtained by setting $\frac{\partial{f_\tau(\mathbf{W})}}{\partial{\mathbf{w}_i}} = \mathbf{0}$:
\vspace{-5pt}
\begin{equation}
    \mathbf{w}_i^\mathsf{opt} = \bigg(
    \alpha_\mathsf{u} | \tau_{\mathsf{u}, i}|^2 \boldsymbol{\zeta} \bigg)^{-1}
    \bigg(\!
    \alpha_\mathsf{u} \sqrt{1+\iota_{\mathsf{u}, i}} \sqrt{P_u} \tau_{\mathsf{u}, i}^* {\mathbf{h}}_{\mathsf{u},i, \mathsf{BS}}
   \!  \bigg),
\label{eq:woptimal}
\end{equation}
\vspace{-5pt}
where $\boldsymbol{\zeta}$ is given by
\begin{equation}
\begin{aligned}
    \boldsymbol{\zeta} & = P_u \sum_{p \in \mathcal{I}}  {\mathbf{h}}_{\mathsf{u},p, \mathsf{BS}} {\mathbf{h}}_{\mathsf{u},p, \mathsf{BS}}^H  + \mathbf{H}_\mathsf{SI}  \sum_{k \in \mathcal{K}} \mathbf{p}_k \mathbf{p}_k^H \mathbf{H}_\mathsf{SI}^H \\
    & + \mathbf{G}^\top (\mathbf{\Phi} - \mathbf{I})\mathbf{G} \sum_{k \in \mathcal{K}} \mathbf{p}_k \mathbf{p}_k^H \mathbf{G}^H (\mathbf{\Phi} - \mathbf{I})^H \mathbf{G}^* \\
    & + 2 \Re{\mathbf{G}^\top (\mathbf{\Phi} - \mathbf{I})\mathbf{G} \sum_{k \in \mathcal{K}} \mathbf{p}_k \mathbf{p}_k^H \mathbf{H}_\mathsf{SI}^H}
    + \sigma^2.
\end{aligned}
\end{equation}

Finally, $\mathbf{W}$ is normalized (\ie $\mathbf{W}/\| \mathbf{W}\|_F$) to satisfy the receiver side power constraint.

\bpara{BD-RIS Scattering Matrix: $\mathbf{\Phi}$.} With the given $\mathbf{P}, \mathbf{W}, \boldsymbol{\iota}, \boldsymbol{\tau}$, we extract the terms w.r.t $\mathbf{\Phi}$ given by
\begin{equation}
    \begin{aligned}
        &f_\tau(\mathbf{\Phi}) = \alpha_\mathsf{d} \sum_{k \in \mathcal{K}} 2 \sqrt{1+\iota_{\mathsf{d}, k}} \Re{\tau_{\mathsf{d}, k}^* \mathbf{h}_{\mathsf{ref, d}, k}^\top \mathbf{\Phi} \mathbf{G} \mathbf{p}_k}  \\
        &  +  \alpha_\mathsf{u} \sqrt{P_u} \sum_{i \in \mathcal{I}} 2 \sqrt{1+\iota_{\mathsf{u}, i}} \Re{\tau_{\mathsf{u}, i}^* \mathbf{w}_i^H \mathbf{G}^\top  \mathbf{\Phi} \mathbf{h}_{\mathsf{ref}, \mathsf{u}, i}}  \\
        & - \alpha_\mathsf{d} \sum_{k \in \mathcal{K}} |\tau_{\mathsf{d}, k}|^2 \bigg(
        \sum_{j \in \mathcal{K}} |{\mathbf{h}}_{\mathsf{d}, k}^\top \mathbf{p}_j|^2 + P_u \sum_{i \in \mathcal{I}} |\mathbf{h}_{\mathsf{u}, i, k}|^2
        \bigg)  \\
        & - \alpha_\mathsf{u} \!\! \sum_{i \in \mathcal{I}}\! |\tau_{\mathsf{u}, i}|^2 \!  \bigg( \! \!
        P_u  \sum_{p \in \mathcal{I}} |\mathbf{w}_i^H {\mathbf{h}}_{\mathsf{u},p, \mathsf{BS}}|^2   + \sum_{k \in \mathcal{K}} |\mathbf{w}_i^H \mathbf{H}_\mathsf{SI}\mathbf{p}_k \\
        &  \qquad + \mathbf{w}_i^H \mathbf{G}^\top (\mathbf{\Phi} \! - \!\mathbf{I}) \mathbf{G} \mathbf{p}_k|^2 \!
        \bigg)\\
        & = 2 \alpha_\mathsf{d} \Re{ \Tr({\mathbf{C}_1}  \mathbf{\Phi})} + 2 \alpha_\mathsf{u} \sqrt{P_u} \Re{\Tr({\mathbf{C}_2}\mathbf{\Phi} )}\\
        & \quad - \alpha_\mathsf{d} \Tr({\mathbf{B}_1} \mathbf{\Phi} {\mathbf{A}_1} \mathbf{\Phi}^H ) + 2 \alpha_\mathsf{d} \Re{\Tr({\mathbf{F}_1} \mathbf{\Phi})}\\
        & \quad - 2 \alpha_\mathsf{d} \Re{\Tr({\mathbf{D}_1} \mathbf{\Phi})}   - \alpha_\mathsf{d} P_u \Tr({\mathbf{B}_1} \mathbf{\Phi}{\mathbf{B}_2} \mathbf{\Phi}^H) \\
        & \quad + 2 \alpha_\mathsf{d} P_u \Re{\Tr({\mathbf{J}_1} \mathbf{\Phi})} -  2 \alpha_\mathsf{d} P_u \Re{\Tr({\mathbf{D}_2} \mathbf{\Phi})}  \\
        & \quad - \alpha_\mathsf{u} P_u \Tr({\mathbf{A}_2} \mathbf{\Phi} {\mathbf{B}_2}  \mathbf{\Phi}^H) + 2 \alpha_\mathsf{u}  P_u \Re{\Tr({\mathbf{F}_2} \mathbf{\Phi})} \\ 
        & \quad - 2 \alpha_\mathsf{u}  P_u \Re{\Tr({\mathbf{D}_3} \mathbf{\Phi})}  - \alpha_\mathsf{u} \Tr(  {\mathbf{A}_2} \mathbf{\Phi} {\mathbf{A}_1} \mathbf{\Phi}^H) \\
        & \quad + 2 \alpha_\mathsf{u} \Re{  \Tr({\mathbf{J}_2} \mathbf{\Phi}) }- 2\alpha_\mathsf{u} \Re{\Tr( \mathbf{D} \mathbf{\Phi})},
    \end{aligned}
    \label{eq:ftau_phi}
    \end{equation}
where the fresh notations above are defined as in the Table \ref{tab:1}.
\begin{table}[t]
\caption{Newly Introduced Notations}
\resizebox{0.5\textwidth}{!}{
\renewcommand{\arraystretch}{1.6}
\begin{tabular}{|l|l|}
\hline
$\mathbf{A}_1 = \sum_{j \in \mathcal{K}} \mathbf{G} \mathbf{p}_j \mathbf{p}_j^H \mathbf{G}^H$ & $\mathbf{F}_1 = \mathbf{A}_1 \mathbf{I}  \mathbf{B}_1$ \\ \hline
$\mathbf{A}_2 = \sum_{i \in \mathcal{I}} |\tau_{\mathsf{u}, i}|^2 \mathbf{G}^* \mathbf{w}_i \mathbf{w}_i^H \mathbf{G}^\top$  & $\mathbf{F}_2 = \mathbf{B}_2 \mathbf{I}  \mathbf{A}_2$ \\ \hline
$ \mathbf{B}_1 = \sum_{k \in \mathcal{K}} |\tau_{\mathsf{d}, k}|^2 \mathbf{h}_{\mathsf{ref, d}, k}^* \mathbf{h}_{\mathsf{ref, d}, k}^\top$  & $\mathbf{J}_1 = \mathbf{B}_2 \mathbf{I}  \mathbf{B}_1$ 
\\ \hline
$\mathbf{B}_2 = \sum_{p \in \mathcal{I}} \mathbf{h}_{\mathsf{ref, u},p} \mathbf{h}_{\mathsf{ref, u},p}^H$  & $\mathbf{J}_2 = \mathbf{A}_1 \mathbf{I}  \mathbf{A}_2$ \\ \hline
$\mathbf{C}_1 = \sum_{k \in \mathcal{K}} \sqrt{1+\iota_{\mathsf{d}, k}} \tau_{\mathsf{d}, k}^* \mathbf{G} \mathbf{p}_k \mathbf{h}_{\mathsf{ref, d}, k}^\top$ 
& 
$\mathbf{C}_2 = \sum_{i \in \mathcal{I}} \sqrt{1+\iota_{\mathsf{u}, i}} \tau_{\mathsf{u}, i}^* \mathbf{h}_{\mathsf{ref},\mathsf{u},i} \mathbf{w}_i^H \mathbf{G}^\top$ 
\\ 
\hline
\multicolumn{2}{|l|}{$\mathbf{D}_1 = \mathbf{G} \sum_{j \in \mathcal{K}} \mathbf{p}_j \mathbf{p}_j^H \sum_{k \in \mathcal{K}} |\tau_{\mathsf{d}, k}|^2 \mathbf{h}_{\mathsf{dir, d}, k}^* \mathbf{h}_{\mathsf{ref, d}, k}^\top  $}\\
\hline
\multicolumn{2}{|l|}{$\mathbf{D}_2 =  \sum_{i \in \mathcal{I}} \mathbf{h}_{\mathsf{ref, u}, i} (\sum_{k \in \mathcal{K}} |\tau_{\mathsf{d}, k}|^2 \mathbf{h}_{\mathsf{dir, u}, i, k} \mathbf{h}_{\mathsf{ref, d}, k}^\top)$}\\
\hline
\multicolumn{2}{|l|}{$\mathbf{D}_3 =  \sum_{p \in \mathcal{I}} \mathbf{h}_{\mathsf{ref, u}, p} \mathbf{h}^H_{\mathsf{dir, u}, p, \mathsf{BS}} \sum_{i \in \mathcal{I}} |\tau_{\mathsf{u}, i}|^2 \mathbf{w}_i \mathbf{w}_i^H \mathbf{G}^\top $}\\
\hline
\multicolumn{2}{|l|}{$\mathbf{D} = \mathbf{G} \sum_{j \in \mathcal{K}} \mathbf{p}_j \mathbf{p}_j^H \mathbf{H}_\mathsf{SI}^H \sum_{i \in \mathcal{I}} |\tau_{\mathsf{u}, i}|^2 \mathbf{w}_i \mathbf{w}_i^H \mathbf{G}^\top  $}\\
\hline
\end{tabular}
\label{tab:1}
}
\end{table}
% Taking the group-connected reciprocal BD-RIS as the example, the sub-problem w.r.t $\mathbf{\Phi}$ is formulated as 
% \begin{maxi!}|s|[2]                   % mini! = minimize 
%     {\mathbf{\Phi}}                               % optimization variable
%     {f_\tau(\mathbf{\Phi}) \label{eq:op6}}   % objective function and label
%     {\label{eq:p6}}             % label for opt problem
%     {\mathcal{P}6:} 
%     % optimization result
%     {}
%     \addConstraint{\mathbf{\Phi}_g^H \mathbf{\Phi}_g = \mathbf{I},}{\quad  \forall g \in \mathcal{G} \label{eq:op6c1},}
%     \addConstraint{\mathbf{\Phi}_g = \mathbf{\Phi}_g^\top,}{\quad \forall g \in \mathcal{G} \label{eq:op6c2},}
%     \addConstraint{\mathbf{\Phi} = \operatorname{blkdiag}(\mathbf{\Phi}_1, \cdots, \mathbf{\Phi}_G).}{\label{eq:op6c3}}
% \end{maxi!}
Since the group-connected BD-RIS includes single- and fully-connected BD-RIS as special cases, we focus on the following general sub-problem w.r.t $\mathbf{\Phi}$, which is formulated as 
\begin{maxi!}|s|[2]                   % mini! = minimize 
    {\mathbf{\Phi}}                               % optimization variable
    {f_\tau(\mathbf{\Phi}) \label{eq:op6}}   % objective function and label
    {\label{eq:p6}}             % label for opt problem
    {\mathcal{P}6:} 
    % optimization result
    {}
    \addConstraint{\mathbf{\Phi}\in\mathcal{R}_i, \, i \in\{1,2\},}{\label{eq:op6c2}}
    \addConstraint{\mathbf{\Phi}}{\in \mathcal{S}_2 \label{eq:op6c1}.}
    % \addConstraint{\mathbf{\Phi} = \operatorname{blkdiag}(\mathbf{\Phi}_1, \cdots, \mathbf{\Phi}_G)}{\label{eq:op6c3},}
\end{maxi!}
Subsequently, we provide a general solution to design the scattering matrix, $\mathbf{\Phi}$, for non-reciprocal and reciprocal BD-RISs. 
In the following subsections, we will separately tackle the two difficult constraints: the self-orthogonality (unitary) constraint in \eqref{eq:op6c1} and the symmetry constraint in \eqref{eq:op6c2}. 

% There are two difficulties in solving $\mathcal{P}6$; the self-orthogonality constraint $\mathbf{\Phi}_g^H \mathbf{\Phi}_g = \mathbf{I}$  and the symmetry constraint $\mathbf{\Phi}_g = \mathbf{\Phi}_g^\top$ \eqref{eq:op6c2}. 

\begin{algorithm}[t]
	\caption{PDD Algorithm for Updating BD-RIS Scattering Matrix $\mathbf{\Phi}$}
	\label{alg:alg2}
	\KwIn{$\mathbf{P}, \mathbf{W}, \boldsymbol{\iota}, \boldsymbol{\tau},  \mathbf{H}_{\mathsf{d}}, \mathbf{H}_{\mathsf{u, BS}}, \mathbf{H}_{\mathsf{u, DL}}, \mathbf{H}_\mathsf{SI}, \mathbf{G}$.}  
	\KwOut{$\mathbf{\Phi}^\mathsf{opt}$.} 
	% \BlankLine
	Initialize $\{\mathbf{\Phi}_g\}, \{\mathbf{\Psi}_g\}, \{\mathbf{\Lambda}_g\}, \rho, t_\mathsf{inner}=t_\mathsf{outer}=1$.
 
    \For{$g \gets 1$ \KwTo $G$}{
        \While{\textnormal{$\| \mathbf{\Phi}_g - \mathbf{\Psi}_g \|_\infty > \varepsilon$ \textbf{\&}} $ t_\mathsf{outer}<t_\mathsf{outer\, max}$ }{
        \While{\textnormal{no convergence of objective function \eqref{eq:op7} \textbf{\&}} $ t_\mathsf{inner}<t_\mathsf{inner\, max}$ }{
        Update $ \mathbf{\Phi}_g $ by \eqref{eq:phi}.\\
        Update $ \mathbf{\Psi}_g $ by \eqref{eq:phi_copy}.\\
        $t_\mathsf{inner} = t_\mathsf{inner} + 1$.
	}
    \uIf{$\| \mathbf{\Phi}_g - \mathbf{\Psi}_g \|_\infty < \epsilon$}{
        $\mathbf{\Lambda}_g = \mathbf{\Lambda}_g + \rho^{-1} (\mathbf{\Phi}_g - \mathbf{\Psi}_g)$.
    }
    \Else{
        $\rho = c\rho$.
    }
    $t_\mathsf{outer} = t_\mathsf{outer} + 1$.
	}	
    }
	\KwRet{ $\mathbf{\Phi}^\mathsf{opt} = \operatorname{blkdiag}(\mathbf{\Phi}^\mathsf{opt}_1, \cdots, \mathbf{\Phi}^\mathsf{opt}_G)$.}
\end{algorithm}
\subsection{PDD Method to Decouple \eqref{eq:op6c1} and \eqref{eq:op6c2}}
Inspired by research adopting the PDD method \cite{shi2020penalty, zhou2023optimizing}, we adopt it to decouple constraints  \eqref{eq:op6c1} and \eqref{eq:op6c2}.
Specifically, the PDD method is a two-loop iterative algorithm \cite{shi2020penalty}, where the inner loop solves the augmented Lagrangian problem using the BCD method. In our case, each $\mathbf{\Phi}_g, \forall g \in \mathcal{G}$ is designed by iteration. In each iteration, with the fixed $ \mathbf{\Lambda}_g $ and $\rho$, the PDD framework alternately updates the $ \mathbf{\Phi}_g $ and $ \mathbf{\Psi}_g $ until the minimization of the objective function. In the outer loop, the PDD framework selectively updates $\{ \mathbf{\Lambda}_g \}$ and $\rho$ until convergence (\ie $\| \mathbf{\Phi} - \mathbf{\Psi} \|_\infty \leq \varepsilon$). The PDD algorithm for updating $\mathbf{\Phi}$ is summarized in Algorithm \ref{alg:alg2}. 
Specifically, we introduce a copy $\{\mathbf{\Psi}_g\}$ of $\{\mathbf{\Phi}_g\}$ with an equality constraint $\mathbf{\Psi}_g = \mathbf{\Phi}_g, \forall g \in \mathcal{G}$. The reformulated problem is given by
% \begin{mini!}|s|[2]                   % mini! = minimize 
%     {\mathbf{\Phi}, \mathbf{\Psi}}                               % optimization variable
%     {f_\tau(\mathbf{\Phi}) \label{eq:op70}}   % objective function and label
%     {\label{eq:p70}}             % label for opt problem
%     {\mathcal{P}7:} 
%     % optimization result
%     {}
%     \addConstraint{\mathbf{\Psi}_g^H \mathbf{\Psi}_g = \mathbf{I},}{\quad  \forall g \in \mathcal{G} \label{eq:op70c1},}
%     \addConstraint{\mathbf{\Phi}_g = \mathbf{\Psi}_g,}{\quad  \forall g \in \mathcal{G} \label{eq:op70c2},}
%     \addConstraint{\eqref{eq:op6c2}.}{\nonumber}
% \end{mini!}
\begin{maxi!}|s|[2]                   % mini! = minimize 
    {\mathbf{\Phi}, \mathbf{\Psi}}                               % optimization variable
    {f_\tau(\mathbf{\Phi}) \label{eq:op70}}   % objective function and label
    {\label{eq:p70}}             % label for opt problem
    {\mathcal{P}7:} 
    % optimization result
    {}
    \addConstraint{\mathbf{\Psi}_g^H \mathbf{\Psi}_g = \mathbf{I},}{\quad  \forall g \in \mathcal{G} \label{eq:op70c1},}
    \addConstraint{\mathbf{\Phi}_g = \mathbf{\Psi}_g,}{\quad  \forall g \in \mathcal{G} \label{eq:op70c2},}
    \addConstraint{\eqref{eq:op6c2}.}{\nonumber}
\end{maxi!}
Subsequently, the PDD framework \cite{shi2020penalty} is adopted and we formulate the augmented Lagrangian function by penalizing the equality constraint \eqref{eq:op70c2}, expressed by
\begin{equation}
\begin{aligned}
\label{eq:obj_pdd}
    L( \{ \!\mathbf{\Phi}_g \!\}, \{\!\mathbf{\Psi}_g\! \}, &\{\!\mathbf{\Lambda}_g \!\},\rho) \! =\! - \!f_\tau(\mathbf{\Phi}) \!+  \!\frac{1}{2 \rho}  \!\sum_{g \in \mathcal{G}}\! \|\mathbf{\Phi}_g \! - \! \mathbf{\Psi}_g \|^2 \\
    & + \sum_{g \in \mathcal{G}} \Re{\Tr(\mathbf{\Lambda}_g^H(\mathbf{\Phi}_g - \mathbf{\Psi}_g))}, 
\end{aligned}
\end{equation}
where $\mathbf{\Lambda}_g \in \mathbb{C}^{M \times M}, \forall g \in \mathcal{G}$, and $1/\rho$ are the Lagrangian dual variable and penalty coefficient in the PDD framework, respectively. Thus, $\mathcal{P}$7 is reformulated as 
\begin{mini!}|s|[2]                   % mini! = minimize 
    {\mathbf{\Phi}, \mathbf{\Psi}}                               % optimization variable
    {L( \{ \mathbf{\Phi}_g \}, \{\mathbf{\Psi}_g \}, \{\mathbf{\Lambda}_g \},\rho\,)  \label{eq:op7}}   % objective function and label
    {\label{eq:p7}}             % label for opt problem
    {\mathcal{P}8:} 
    % optimization result
    {}
    \addConstraint{\eqref{eq:op6c2}, \eqref{eq:op70c1}.}{\nonumber}
\end{mini!}
The following sections explain the details for updating the inner and outer loops in the PDD method.

\subsection{Inner Loop}
% The PDD method is a two-loop iterative algorithm \cite{shi2020penalty}, where the inner loop solves the augmented Lagrangian problem using the BCD method. In our case, each $\mathbf{\Phi}_g, \forall g \in \mathcal{G}$ is designed by iteration. In each iteration, with the fixed $ \mathbf{\Lambda}_g $ and $\rho$, the PDD framework alternately updates the $ \mathbf{\Phi}_g $ and $ \mathbf{\Psi}_g $ until the minimization of the objective function. In the outer loop, the PDD framework selectively updates $\{ \mathbf{\Lambda}_g \}$ and $\rho$ until convergence (\ie $\| \mathbf{\Phi} - \mathbf{\Psi} \|_\infty \leq \varepsilon$). The PDD algorithm for updating $\mathbf{\Phi}$ is summarized in Algorithm \ref{alg:alg2}. The following sections explain the details for updating the inner and outer loops in the PDD method.

In this section, we first explain the linear reformulation of the symmetry constraint. Based on this, the detailed update process is then provided.

\subsubsection{Linear Reformulation of Symmetry Constraint}
The main idea of addressing symmetry constraint is to design the neccessary elements in the matrix: $i)$ all elements of $\mathbf{\Phi}_g$ for non-reciprocal BD-RIS, and $ii)$ the diagonal and lower-triangular elements of $\mathbf{\Phi}_g$ for reciprocal BD-RIS. We extract these required elements and store them in a vector $\boldsymbol{\varphi}_g$ in a column-by-column order. To recover $\boldsymbol{\phi}_g = \operatorname{vec}\left(\boldsymbol{\Phi}_g\right)$ from $\boldsymbol{\varphi}_g$, a permutation matrix is introduced. The permutation process is represented by
\begin{equation}
    \boldsymbol{\phi}_g \triangleq \operatorname{vec}\left(\boldsymbol{\Phi}_g\right)=\mathbf{K}_{g} \boldsymbol{\varphi}_g, \quad \forall g \in \mathcal{G}.
\label{eq:mat2vec}
\end{equation}
The permutation matrix $\mathbf{K}_{g}$ has different forms:
% for the two cases:
\begin{enumerate}[leftmargin = *, label={\roman*)}]
    \item non-reciprocal BD-RIS: $\mathbf{K}_g \in \mathbb{B}^{M_g^2 \times M_g^2}$ and $\mathbf{K}_g  = \mathbf{I}$, 

    \item reciprocal BD-RIS: $\mathbf{K}_g \in \mathbb{B}^{M_g^2 \times \frac{M_g(M_g+1)}{2}}$, and 
\end{enumerate}
\begin{equation}
    \mathbf{K}_g(i,j) \! =  \! \left \{ 
    \begin{aligned}
        \begin{aligned}
            1&, 
            \begin{aligned}
                \text{if} \,\, \{ i &\!= \!(p\!-\!1)M_g+q \,\, \text{or} \,\,  (q\!-\!1)M_g\!+\!p \}, \, \text{and}\\
                j & = \frac{(q-1)(2M_g-q+2)}{2}+(p-q+1)\\
            \end{aligned}\\
            % \text{if} \,\, i = (p-1)M_g+q \,\, \text{or} \,\,  (q-1)M_g+p\\
            % & \quad j = \frac{(q-1)(2M_g-q+2)}{2}+(p-q+1)\\
            0&, \, \text{otherwise},
        \end{aligned}
    \end{aligned}
     \right.
\end{equation}
where $p$ and $q$ are the row and column indices of the diagonal and lower-triangular elements in $\mathbf{\Phi}_g$, \ie $\mathbf{\Phi}_g(p,g)$ with $1 \leq p \leq M_g$ and $1 \leq q \leq p$.

To map the group vector $\boldsymbol{\phi}_g$ to its corresponding position in the $\boldsymbol{\phi}$, we introduce reshaping matrices $\{ \mathbf{R}_g \}$, where $\mathbf{R}_g \in \mathbb{B}^{M^2 \times M_g^2}$. The mapping process is express as
\begin{equation}
\boldsymbol{\phi} \triangleq \operatorname{vec}(\boldsymbol{\Phi})=\sum_{g \in \mathcal{G}} \mathbf{R}_g \boldsymbol{\phi}_g,
\end{equation}
where $\mathbf{R}_g, \forall g \in \mathcal{G}$ is expressed by
\begin{equation}
    \mathbf{R}_g(i,j) = \left \{ 
    \begin{aligned}
        \begin{aligned}
            1&, 
            \begin{aligned}
                \text{if} \,\, i &= (M_g(g-1)+m-1)M \\
                                 & \quad + (M_g(g-1)+n), \, \text{and}\\
                               j &= (m-1)M_g + n\\
            \end{aligned}\\
            % \text{if} \,\, i = (p-1)M_g+q \,\, \text{or} \,\,  (q-1)M_g+p\\
            % & \quad j = \frac{(q-1)(2M_g-q+2)}{2}+(p-q+1)\\
            0&, \, \text{otherwise},
        \end{aligned}
    \end{aligned}
     \right.
\end{equation}
where $1 \leq m \leq M_g$ and $1 \leq n \leq M_g$ are iteration indices.
After preprocessing the constraints, we can subsequently utilize the PDD method to obtain the optimal solution for the scattering matrix.

\subsubsection{$\mathbf{\Phi}_g$ in the inner loop}
We first extract the terms w.r.t $\mathbf{\Phi}_g$ in the objective function and constraints. 
With the above transformations and the following equations,
\begin{equation}
\operatorname{Tr}(\mathbf{C}\boldsymbol{\Phi})\!=\!\operatorname{vec}^\top\left(\mathbf{C}^\top\right) \boldsymbol{\phi}\!=\!\sum_{g \in \mathcal{G}} \operatorname{vec}^\top\left(\mathbf{C}^\top\right) \mathbf{R}_g \mathbf{K}_{g} \boldsymbol{\varphi}_g,
\end{equation}
\begin{equation}
\begin{aligned}
    \operatorname{Tr}&\left(\mathbf{B} \boldsymbol{\Phi} \mathbf{A} \boldsymbol{\Phi}^H\right)\!=\!\boldsymbol{\phi}^H\left(\mathbf{A}^\top \otimes \mathbf{B}\right) \boldsymbol{\phi} \\
    &=\big(\sum_{g \in \mathcal{G}} \mathbf{R}_g \mathbf{K}_{g} \boldsymbol{\varphi}_g\big)^H   \!\! \left(\mathbf{A}^\top \! \! \otimes \! \mathbf{B}\right) \!\big(\sum_{g \in \mathcal{G}} \mathbf{R}_g \mathbf{K}_{g} \boldsymbol{\varphi}_g \big),
\end{aligned}
\end{equation}
the optimization problem w.r.t each $\mathbf{\Phi}_g $ can be simplified to 
\begin{mini!}|s|[2]                   % mini! = minimize 
    {\boldsymbol{\varphi}_g}                               % optimization variable
    {L( \boldsymbol{\varphi}_g ) = \boldsymbol{\varphi}_g^H \boldsymbol{\Delta} \boldsymbol{\varphi}_g - 2 \Re{\boldsymbol{\varphi}_g^H \boldsymbol{\delta}}. \label{eq:op8}}   % objective function and label
    {\label{eq:p8}}             % label for opt problem
    {\mathcal{P}9:} 
    % optimization result
    {}
\end{mini!}
We have $\boldsymbol{\psi}_g \triangleq \operatorname{vec}(\boldsymbol{\Psi}_g)$, $\lambda_g \triangleq \operatorname{vec}(\boldsymbol{\Lambda}_g)$, and 
\begin{equation}
\begin{aligned}
    & \boldsymbol{\Delta}  = \alpha_\mathsf{d} \mathbf{K}_{g}^H \mathbf{R}_g^H \big(
    \mathbf{A}_1^\top \otimes \mathbf{B}_1 + P_u \mathbf{B}_2^\top \otimes \mathbf{B}_1 \big) \mathbf{R}_g \mathbf{K}_{g}\\
    & + \alpha_\mathsf{u} \mathbf{K}_{g}^H \mathbf{R}_g^H \big(
    P_u \mathbf{B}_2^\top \otimes \mathbf{A}_2 + \mathbf{A}_1^\top \otimes \mathbf{A}_2 \big) \mathbf{R}_g \mathbf{K}_{g} \\
    & + \frac{1}{2 \rho} \mathbf{K}_{g}^H \mathbf{K}_{g},
\end{aligned}
\end{equation}
\vspace{-10pt}
\begin{equation}
\begin{aligned}
    & \boldsymbol{\delta} =  \alpha_\mathsf{d} \mathbf{K}_{g}^H \mathbf{R}_g^H \bigg(
    \operatorname{vec}^*(\mathbf{C}_1^\top) + \operatorname{vec}^*(\mathbf{F}_1^\top) - \operatorname{vec}^*(\mathbf{D}_1^\top)\\
    & + \! P_u \operatorname{vec}^*(\mathbf{J}_1^\top) \! - \! P_u \operatorname{vec}^*(\mathbf{D}_2^\top)\bigg) \! + \! \alpha_\mathsf{u}  \mathbf{K}_{g}^H \mathbf{R}_g^H \bigg(
    \sqrt{P_u} \operatorname{vec}^*(\mathbf{C}_2^\top) \! \\
    &  + \! P_u \operatorname{vec}^*(\mathbf{F}_2^\top) - P_u \operatorname{vec}^*(\mathbf{D}_3^\top) \! + \operatorname{vec}^*(\mathbf{J}_2^\top) - \operatorname{vec}^*(\mathbf{D}^\top)
    \bigg) \\
    &+ \frac{1}{2\rho} \mathbf{K}_{g}^H \mathbf{K}_{g} \boldsymbol{\psi}_g - \frac{1}{2} \mathbf{K}_{g}^H \boldsymbol{\lambda}_g.
\end{aligned}
\end{equation}

Since $\mathcal{P}9$ is an unconstrained convex problem, we can take the derivative of $L(\boldsymbol{\varphi}_g )$ w.r.t $\boldsymbol{\varphi}_g$ and set $\frac{\partial{L( \boldsymbol{\varphi}_g ) }}{\partial{\boldsymbol{\varphi}_g}} = \mathbf{0}$ to obtain optimal $\boldsymbol{\varphi}_g^*$, which is given by
\begin{equation}
    \boldsymbol{\varphi}_g^* = \boldsymbol{\Delta}^{-1} \boldsymbol{\delta}.
    \label{eq:phi}
\end{equation}

\subsubsection{$\mathbf{\Psi}_g$ in the inner loop} With other variables determined, the optimization problem relates to $\mathbf{\Psi}_g$ is given by
\begin{mini!}|s|[2]                   % mini! = minimize 
    {\boldsymbol{\Psi}_g}                               % optimization variable
    {\| \boldsymbol{\Psi}_g - (\rho \boldsymbol{\Lambda}_g + \boldsymbol{\Phi_g})  \|^2_F, \label{eq:op9}}   % objective function and label
    {\label{eq:p9}}             % label for opt problem
    {\mathcal{P}10:} 
    % optimization result
    {}
    \addConstraint{\mathbf{\Psi}_g^H \mathbf{\Psi}_g = \mathbf{I}.}{\label{eq:op9c1}}
\end{mini!}
This corresponds to the orthogonal procrustes problem, which has a close-form solution \cite{gibson1962least, manton2002optimization, gloub1996matrix, zhou2023optimizing}, expressed as 
\begin{equation}
\boldsymbol{\Psi}_g^{\mathsf{opt}}=\mathbf{U}_g  \mathbf{V}_g^H,
\label{eq:phi_copy}
\end{equation}
where the unitary matrices $\mathbf{U}_g$, and $\mathbf{V}_g$ are obtained by singular vector decomposition (SVD) of $\rho \boldsymbol{\Lambda}_g + \boldsymbol{\Phi_g}$.

\subsection{Outer loop} When the convergence of the inner loop is achieved, the dual variable $ \mathbf{\Lambda}_g $ and penalty coefficient $\rho^{-1}$ are selectively updated. If $\| \mathbf{\Phi}_g - \mathbf{\Psi}_g \|_\infty < \epsilon$, then the $\mathbf{\Lambda}_g$ is updated by
\begin{equation}
    \mathbf{\Lambda}_g = \mathbf{\Lambda}_g + \rho^{-1} (\mathbf{\Phi}_g - \mathbf{\Psi}_g).
\end{equation}
Otherwise, $\| \mathbf{\Phi}_g - \mathbf{\Psi}_g \|_\infty > \epsilon$ means the infinity distance between these two matrices are still large, thus the $\rho$ is updated to force the equality between $\mathbf{\Phi}_g$ and $\mathbf{\Psi}_g$, given by $\rho = c\rho$,
% \begin{equation}
%     \rho = c\rho,
% \end{equation}
where penalty parameter $c \in (0,1)$. After each $\mathbf{\Phi}_g^\mathsf{opt}$ is obtained, we reconstruct the $\mathbf{\Phi}^\mathsf{opt}$ by mapping each $\mathbf{\Phi}_g^\mathsf{opt}$ to the corresponding position, \ie $\mathbf{\Phi}^\mathsf{opt} = \operatorname{blkdiag}(\mathbf{\Phi}_1, \cdots, \mathbf{\Phi}_G)$.

\bpara{Complexity Analysis.}
We briefly analyze the computational complexity of the proposed algorithm in this section. Within the BCD framework, we iteratively update the blocks.  In each iteration, updating $\boldsymbol{\iota}_{\mathsf{d}}$, and $\boldsymbol{\tau}_{\mathsf{d}}$ require $\mathcal{O}(K^2 M^2)$ operations, and $\boldsymbol{\iota}_{\mathsf{u}}$, and $\boldsymbol{\tau}_{\mathsf{u}}$ require $\mathcal{O}(I^2 M^2)$ operations. The step for updating precoder $\mathbf{P}$ has the complexity $\mathcal{O}(K(M^2+I_p N^3))$ due to the matrix inversion and bisection search, where $I_p$ is the iteration number of bisection search. The step for updating combiner $\mathbf{W}$ requires $\mathcal{O}(I N^3)$ operations due to the matrix inversion. The most computationally intensive step in BCD is updating the scattering matrix $\mathbf{\Phi}$ using the PDD based Algorithm \ref{alg:alg2}. The calculations of \eqref{eq:phi} and \eqref{eq:phi_copy} require $\mathcal{O}(M_g^2 M^4)$, primarily due to the Kronecker product and $\mathcal{O}(M_g^3)$ operations due to SVD, respectively. Therefore, the computational complexity of Algorithm \ref{alg:alg2} is $\mathcal{O}(G I_1 I_2 (M_g^2 M^4))$, where $G$ is the total number of groups, $I_1$ is the number of PDD outer iteration, and $I_2$ is the number of PDD inner iterations in the PDD. Considering the BCD framework in Algorithm \ref{alg:alg1}, the total computational complexity of the entire optimization algorithm is $\mathcal{O}(I_3 G I_1 I_2 (M_g^2 M^4))$, where $I_3$ is the number of BCD iteration until convergence.

\section{Numerical evaluation}
\label{sec:simu}
The objective of this section is to illustrate the superiority of the non-reciprocal BD-RIS over the reciprocal BD-RIS and conventional RIS (\ie D-RIS) in FD communication scenarios. Starting from the investigation of the convergence behaviours, we demonstrate that the proposed algorithm guarantees convergence. Subsequently, we utilize the proposed algorithm to solve the FD DL and UL sum-rate maximization problem as formulated in $\mathcal{P}1$. We compare the FD DL and UL sum-rates of $\textit{i}$) non-reciprocal BD-RIS, $\textit{ii}$) reciprocal BD-RISs, and $\textit{iii}$) D-RIS versus RIS elements, locations of devices,  group size in RIS. Additionally, we analyze the impinging and reflected beampattern performances of the three types of RIS and explain why the non-reciprocal BD-RIS achieves the best performance.  By varying the priority parameter, $\alpha_\mathsf{d}$, between DL and UL communications, we observe the trade-offs of the three cases. We consider both cases which are single DL and UL user and multiple DL and UL users. Note that the results for the model without structural scattering can be obtained by removing $- \mathbf{I}$ right after $\mathbf{\Phi}$ in all channel expressions. {In the simulations, we include the direct links in the convergence analysis to demonstrate the general applicability of the proposed algorithm. Direct links are also considered in the sum-rate performance evaluation with respect to the number of RIS elements. These results show that the non-reciprocal BD-RIS still provides performance gains over reciprocal BD-RIS and D-RIS even in the presence of direct links. In the subsequent simulations, to better highlight the benefits of the non-reciprocal BD-RIS, we assume that all direct links are blocked.}
\vspace{-10pt}
\subsection{Simulation Environment}
\begin{figure}[t]
    \centering
    \includegraphics[width = 0.3\textwidth]{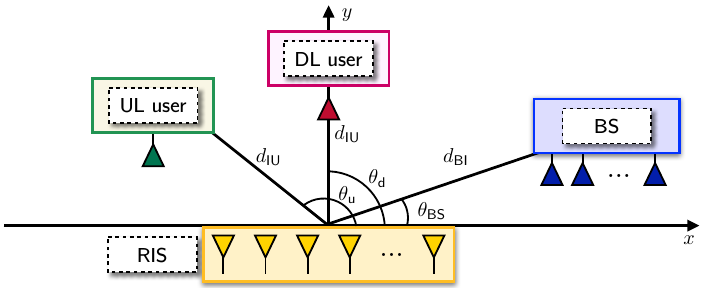}
    \centering
    \caption{2D coordinate system for the RIS-assisted FD system.}
    \label{fig:simu_con}
\end{figure}
The 2D coordinate system for the RIS-assisted FD system is given in \fig{fig:simu_con}. We consider a FD BD-RIS-assisted communication system where the BS is equipped with $N_t$ transmit antenna elements and $N_r$ receive antenna elements. We assume the $N_t=N_r=N$ for simplicity. All DL and UL users are deployed with single antenna. The channel models between BS and RIS, and between RIS and DL/UL users, consist of both large-scale and small-scale fading, consistent with the existing RIS studies \cite{wu2019intelligent, yildirim2020modeling}. The large-scale fading is described by a distance-dependent pathloss model $\mathsf{P L}_i=\zeta_0\left(d_i / d_0\right)^{-\varepsilon_i}, \forall i \in\{\mathsf{BI}, \mathsf{IU}\}$. The $\zeta_0$ denotes the signal attenuation at a reference distance $d_0 = 1$ m, and $d_\mathsf{BI}$ and $d_\mathsf{IU}$ refers to the distance between BS and RIS, and between RIS and DL/UL users, respectively. The path loss exponent is modeled by $\varepsilon_i, \forall i \in\{\mathsf{BI}, \mathsf{IU}\}$. The small-scale fading follows the Rician fading model, characterized by the Rician fading factor $\kappa_i, \forall i \in\{\mathsf{BI}, \mathsf{IU}\}$, which represents the power ratio of line of sight (LoS) component and non-LoS component. {In the simulations, we assume that the direct links are weak as in \cite{zhou2023optimizing}, therefore the BD-RIS can be utilized to intelligently control the wireless environment.} Specifically, we set $\zeta_0=-30$ dB, and the path loss exponent of reflected links for both BS-RIS and RIS-user channels is $\varepsilon_i = 2.2, \forall i \in\{\mathsf{BI}, \mathsf{IU}\}$. The path loss exponent of direct links between BS and DL/UL users, and UL user to DL user is $\varepsilon_\mathsf{dir} = 5$. The distances are set to $d_\mathsf{BI} =30$ m and $d_\mathsf{IU}= 5$ m. We set $\kappa_i = 10$ to capture the strong LoS component. The $K$ DL users and $I$ UL users are randomly located near the RIS with the same distance. The transmit power at the BS and UL user are $P_d = 20$ dBm and $P_u = 20$ dBm, respectively. The noise power at BS and DL users is set as $\sigma^2 = -80$ dBm, and we assume that the SI is effectively suppressed via cancellation methods in propagation \cite{everett2016softnull}, analog \cite{debaillie2014analog} and digital domain \cite{liu2024full}. {Additionally, we investigate the impact of SI on the system performance.}
Perfect CSI is assumed to be available at the BS \cite{li_channel_2024}. 

% \begin{figure}[t]
%     \centering
%     \includegraphics[width = 0.3\textwidth]{figure/convergence_0428.pdf}
%     \centering
%     \caption{{Convergence illustration for Algorithm \ref{alg:alg1} with RIS element $M=32$ and direct links. The locations of BS, UL user, and DL user are  $30^\circ$, $75^\circ$ and $150^\circ$, respectively. Tx and Rx antennas numbers $N = 2$.}}
%     \label{fig:convergence}
% \end{figure}
\begin{figure}[t]
    \centering
    \includegraphics[width = 0.4\textwidth]{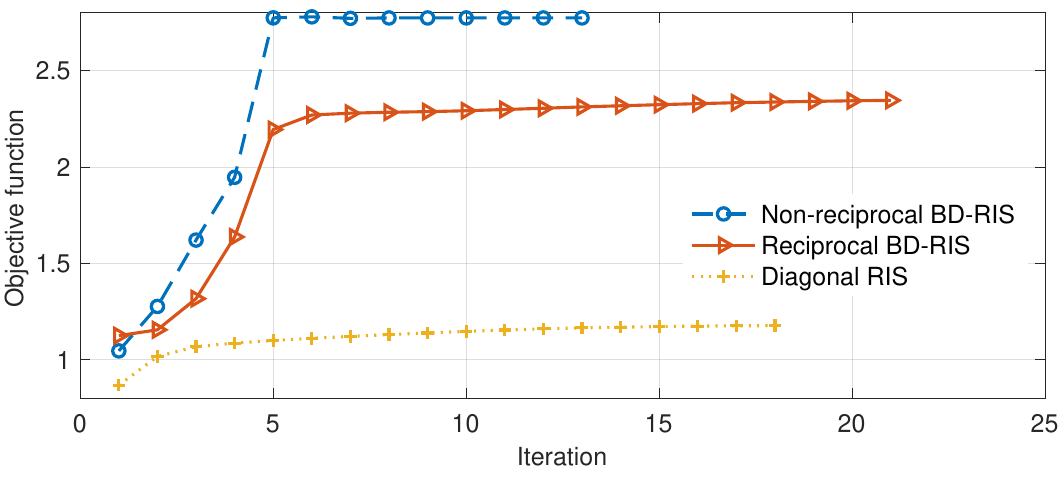}
    \centering
    \caption{{Convergence illustration for Algorithm \ref{alg:alg1} with RIS elements $M = 32$ and direct links. The locations of the BS, UL user, and DL user are $30^\circ$, $75^\circ$, and $150^\circ$, respectively. The numbers of Tx and Rx antennas are $N = 2$, and the numbers of DL and UL users are $K = I = 2$.}}
    \label{fig:convergence}
\end{figure}

\begin{figure}[t]
    \centering
    \includegraphics[width = 0.3\textwidth]{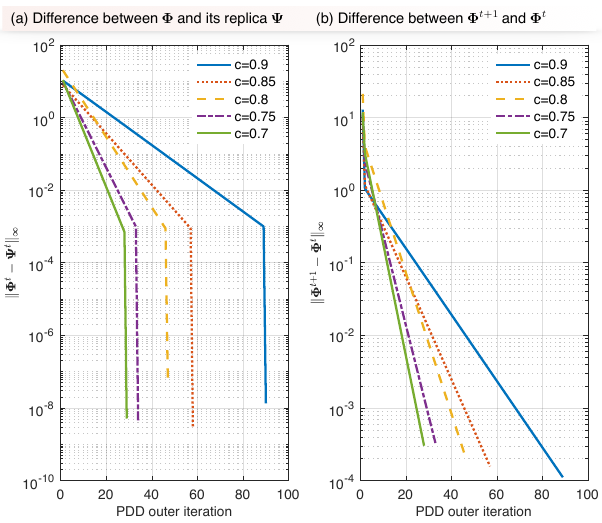}
    \centering
    \caption{{Convergence illustration for Algorithm \ref{alg:alg2} (\ie PDD) with different $c$ and considering direct links. The Tx and Rx  have $N = 2$ antennas, and the UL and DL users are $K=I=2$. The locations of the BS, the co-located UL users, and the co-located DL users are $30^\circ$, $75^\circ$, and $150^\circ$, respectively.}}
    \label{fig:convergence_pdd}
\end{figure}
\vspace{-10pt}
\subsection{Convergence Performance.}
Starting with the convergence analysis, \fig{fig:convergence} (a) shows the values of the objective function \eqref{eq:obj_pdd} versus iterations for fully-connected non-reciprocal and reciprocal BD-RISs, as well as D-RIS, with $N= 2$, $K=I=2$, RIS element number $M=32$, and considering the direct links. It can be observed that the proposed Algorithm \ref{alg:alg1} converges within $25$ iterations for all three types of RIS. Additionally, the fully-connected non-reciprocal BD-RIS achieves the highest value of the objective function. \fig{fig:convergence}. 

% This verifies the effectiveness of the proposed Algorithm \ref{alg:alg1} for group-connected cases. As shown, the objective function values decrease with a reduction in group size $M_g$. 

We also examine the convergence behavior of the PDD method in Algorithm \ref{alg:alg2} for designing the RIS scattering matrix $\mathbf{\Phi}$. \fig{fig:convergence_pdd} (a) and \fig{fig:convergence_pdd} (b) demonstrate the variations of $\mathbf{\Phi}$ and the difference between $\mathbf{\Phi}$ and its replica in the outer loop of the PDD method. It can be observed that Algorithm \ref{alg:alg2} converges within $100$ iterations. The difference in $\mathbf{\Phi}$ between adjacent iterations decreases to a level of $10^{-5}$. Furthermore, the introduced equality constraint \eqref{eq:op70c2} is well satisfied, where $\| \mathbf{\Phi} - \mathbf{\Psi} \|_\infty < 10^{-4}$. Additionally, a smaller penalty parameter $c$ leads to faster convergence to the stopping criteria. A suitable selection for the penalty parameter is $c \in (0.7, 0.9)$. 

% The detailed simulation parameters are provided in Table
% \begin{table}[t]
% \caption{Simulation Parameters}
% \resizebox{0.5\textwidth}{!}{
% % \renewcommand{\arraystretch}{1.5}
% \begin{tabular}{c|c}
% \toprule[1pt]
% \textbf{Parameter} & \textbf{Value} \\ \hline \hline
% $d_\mathsf{BI} $& $30$ m \\ \hline
% $d_\mathsf{IU}$& $5$ m  \\ \hline
% $\zeta_0$ & $-30$ dB \\ \hline
% $\kappa_i, \forall i \in\{\mathsf{BI}, \mathsf{IU}\}$ & $10$ \\ \hline
% Noise Power $\sigma^2$&  $-80$ dBm \\ \hline
% &  \\ \hline
% &  \\ \hline \bottomrule[1pt]
% \end{tabular}
% \label{tab:2}
% }
% \end{table}

\vspace{-10pt}
\subsection{DL and UL Sum-rates over RIS elements}
% \begin{figure}[t]
%     \centering
%     \includegraphics[width = 0.49\textwidth]{figure/ris_1118.pdf}
%     \centering
%     \caption{The sum-rates versus RIS element number of the three types of RIS under (a) no structural scattering, and (b) structural scattering with BS location at $30^\circ$, DL user location at $90^\circ$ and UL user location at $60^\circ$, Tx and Rx antennas numbers $N = 1$.}
%     \label{fig:ris}
% \end{figure}
\begin{figure}[t]
    \centering
    \includegraphics[width = 0.49\textwidth]{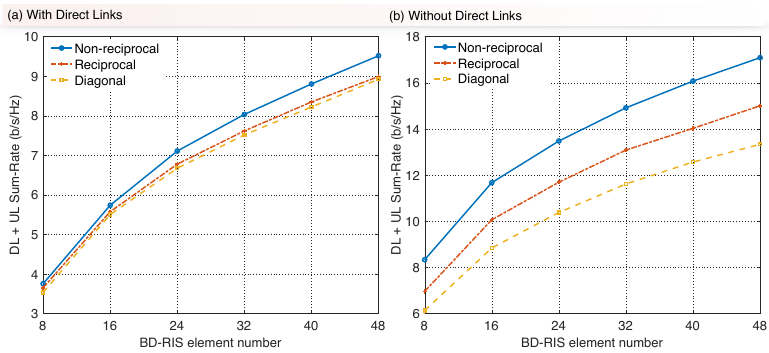}
    \centering
    \caption{{The sum-rates versus the number of RIS elements with structural scattering for the three types of RIS: (a) with direct links, and (b) without direct links. The locations are BS at $30^\circ$, UL user at $75^\circ$, and DL user at $150^\circ$. The Tx and Rx antennas have $N = 1$.}}
    \label{fig:ris}
\end{figure}

In this section, we analyze the DL and UL sum-rate performances to demonstrate the superiority of non-reciprocal BD-RIS in FD communications. Starting with the effect of the number of RIS elements, we fix the locations of the BS, UL user, and DL user at $30^\circ$, $75^\circ$, and $150^\circ$. As shown in \fig{fig:ris} (a), (b), the DL and UL sum-rates increase as the number of RIS elements grows, in both cases with and without direct links. The non-reciprocal BD-RIS consistently achieves the highest sum-rates compared to reciprocal BD-RIS and D-RIS. When considering the direct links, the FD sum-rate performance is lower than the one without direct links. This is due to the direct interference from the UL user to the DL user, which worsens the FD sum-rate. Additionally, the gain of non-reciprocal BD-RIS increases as the number of RIS elements increases. 
\vspace{-10pt}
\subsection{FD DL and UL Sum-rates}
{In this section, we analyze the FD DL and UL sum-rate performances of the three types of RIS versus the locations of the BS, UL user, and DL user. The direct links are assumed to be blocked to better highlight the performance gains of the non-reciprocal BD-RIS.} In addition, we investigate the impinging and reflected beampatterns.
\subsubsection{Sum-rates versus the Location of the UL User}
We start by examining the sum-rate performance relative to the location of the UL user. Specifically, we fix the location of the BS and DL user at $30^\circ$ and $90^\circ$, respectively. Subsequently, we vary the location of the UL user from $0^\circ$ to $180^\circ$. To focus exclusively on observing the effect of different RIS scattering matrix $\mathbf{\Phi}$, we set the number of transmit and receive antenna elements to $N = 1$, the number of DL and UL users to $K=I=1$, and assign $\alpha_\mathsf{d} = \alpha_\mathsf{u} = 0.5$. Since the considered interference terms are different in the DL and UL transmissions as shown in \eqref{eq:dlsignal} and \eqref{eq:rxsignal} respectively, \ie interference from UL users in the DL, and SI and loop interference in the UL, setting $\alpha_\mathsf{d} = \alpha_\mathsf{u}$ does not mean the DL and UL sum-rate perform same. The RIS element number is set to $M=16$. Thus, the optimization problem $\mathcal{P}1$ simplifies to designing only the RIS scattering matrix $\mathbf{\Phi}$ to maximize the FD DL and UL sum-rates.

\begin{figure}[t]
    \centering
    \includegraphics[width = 0.4\textwidth]{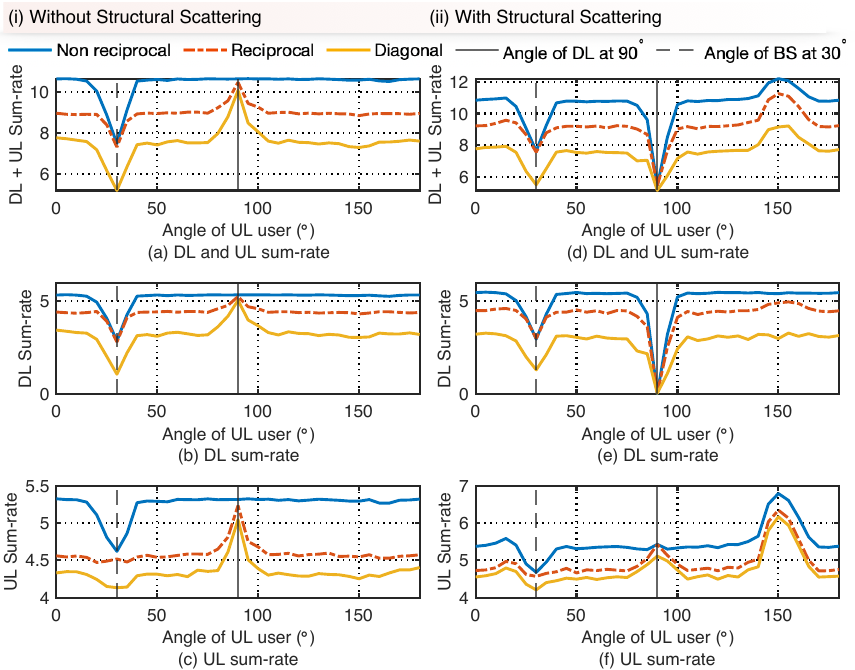}
    \centering
    \caption{DL and UL sum-rates, DL sum-rates, and UL sum-rates for non-reciprocal and reciprocal BD-RISs, and D-RIS with RIS element $M=16$, and fixed BS location at $30^\circ$ and DL user location at $90^\circ$, Tx and Rx antennas numbers $N = 1$, and the number of DL and UL users $K=I=1$. Direct links are assumed to be blocked.}
    \label{fig:angle}
\end{figure}

As shown in \fig{fig:angle} (a), the non-reciprocal BD-RIS achieves the best sum-rates performance except the case that the location of UL user overlaps with the DL user's location at $90^\circ$. This aligns with the explanation in Section \ref{sec:reci}, when $\mathbf{h}_{\mathsf{u},i} = \mathbf{h}_{\mathsf{d}, k}$, reciprocal BD-RIS can maximize both the DL and UL received power, resulting in identical sum-rate performances of the three types of RIS.  In \fig{fig:angle}, the UL sum-rates of reciprocal BD-RIS and D-RIS are larger than their DL sum-rates. This is because the considered interference in UL is smaller than the interference in DL. Specifically, the loop interference in the UL experiences double $d_\mathsf{BI}=30$ m path delay. Its power is smaller than the interference from UL user, which experiences double $d_\mathsf{BI}=5$ m path delay. In contrast, the non-reciprocal BD-RIS can achieve larger received power of signal of interest in the DL compared to other two RISs, thus we can observe the gain in DL. Therefore, the non-reciprocal BD-RIS can support simultaneous DL and UL transmissions.
Furthermore, the sum-rate performances of all three types of RISs deteriorate at $30^\circ$ when the UL user is aligned with the BS. This is because, in the UL SINR expression \eqref{eq:sinr_dlandul}, the ratio of loop interference term to the received signal power approaches 1. Additionally, a performance drop is observed at the supplementary angle of BS at $150^\circ$. Comparing cases that with and without structural scattering in \fig{fig:angle}, we observe that the structural scattering deteriorates the DL performances when DL user and UL user are aligned. This is because the interference term w.r.t structural scattering in DL SINR \eqref{eq:sinr_dlandul}, \ie $|\mathbf{h}_{\mathsf{d}, k}^\top \mathbf{h}_{\mathsf{u}, i}|^2$, which achieves it maximum when $\mathbf{h}_{\mathsf{u},i}$ aligns with $\mathbf{h}_{\mathsf{d}, k}$. 

Additionally, the structural scattering enhances the UL performance at $150^\circ$ because the maximum of structural scattering term is achieved when $\theta_\mathsf{u} = \pi - \theta_\mathsf{BS}$. We then use the UL received power with LoS channel to explain this specular reflection condition. As in \eqref{eq:ul_bound}, the structural scattering term is $|\mathbf{g}^\top \mathbf{h}_{\mathsf{u},i} | $. With the LoS channels 
% \vspace{-5pt}
$\mathbf{g} = \frac{1}{\sqrt{M} } [1, e^{\jmath \pi \cos(\theta_\mathsf{BS})}, \cdots, e^{\jmath \pi (M-1) \cos(\theta_\mathsf{BS})}]^\top$ and $\mathbf{h}_{\mathsf{u},i} = \frac{1}{\sqrt{M} }[1, e^{\jmath \pi \cos(\theta_\mathsf{u})}, \cdots, e^{\jmath \pi (M-1) \cos(\theta_\mathsf{u})}]^\top$.
% \begin{equation}
% \mathbf{g} = \frac{1}{\sqrt{M} } [1, e^{\jmath \pi \cos(\theta_\mathsf{BS})}, \cdots, e^{\jmath \pi (M-1) \cos(\theta_\mathsf{BS})}]^\top,
% \end{equation}
% \vspace{-0pt}
% \begin{equation}
% \mathbf{h}_{\mathsf{u},i} = \frac{1}{\sqrt{M} }[1, e^{\jmath \pi \cos(\theta_\mathsf{u})}, \cdots, e^{\jmath \pi (M-1) \cos(\theta_\mathsf{u})}]^\top.
% \end{equation}
The amplitude of the structural scattering term can be expressed as
$| \mathbf{g}^\top \mathbf{h}_{\mathsf{u},i} | = 1+ | \sum_{n = 1}^{M-1} e^{\jmath n \pi (\cos \theta_\mathsf{BS} + \cos \theta_\mathsf{u} )}  |.$
% \vspace{-10pt}
% \begin{equation}
%     | \mathbf{g}^\top \mathbf{h}_{\mathsf{u},i} | = 1+ | \sum_{n = 1}^{M-1} e^{\jmath n \pi (\cos \theta_\mathsf{BS} + \cos \theta_\mathsf{u} )}  |.
%     \label{eq:strucatual_sca}
% \end{equation}

% When $\theta_\mathsf{BS} + \theta_\mathsf{u} = \pi$, the maximum of \eqref{eq:strucatual_sca} is achieved and has value $M$. 
When $\theta_\mathsf{BS} + \theta_\mathsf{u} = \pi$, the maximum of the structural scattering term is achieved and has value $M$.
This effect on the UL received power also affects the UL sum-rates as in the \fig{fig:angle} (f). Therefore, if the direction of BS is fixed, a sum-rate performance enhancement can theoretically be observed at the angle $\theta_\mathsf{u} = \pi - \theta_\mathsf{BS}$. To further understand why the non-reciprocal BD-RIS achieves superior sum-rate performance, we examine the impinging and reflected beampatterns of the three types of RIS.

\subsubsection{Beampattern}
\label{sssec:beampattern}
\begin{figure}[t]
    \centering
    \includegraphics[width = 0.4\textwidth]{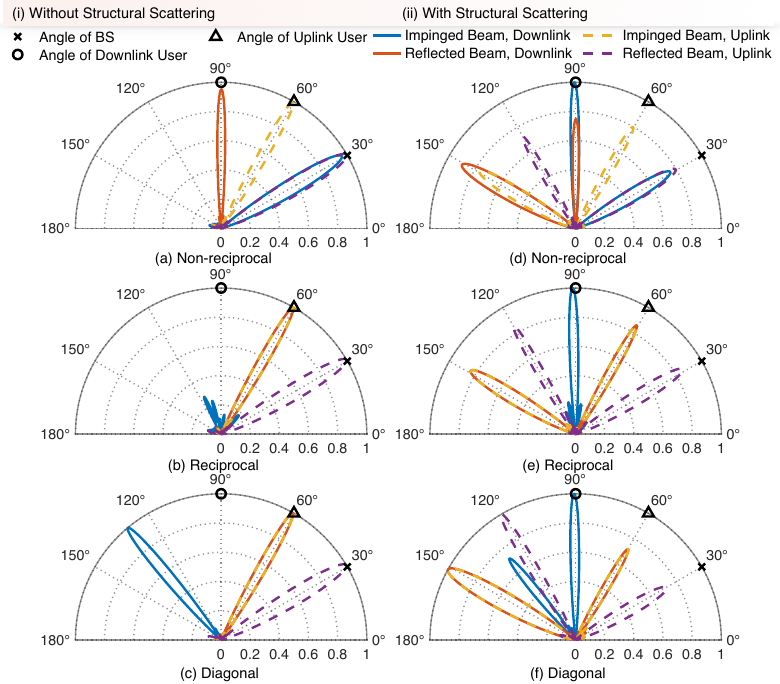}
    \centering
    \caption{The impinging and reflected beampatterns for (i) no structural scattering and (ii) structural scattering. The beampatterns for three types of RIS are shown in (a), (d) non-reciprocal BD-RIS, (b), (e) reciprocal BD-RIS, and (c), (f) D-RIS. The locations of BS, DL user, and UL user are  $30^\circ$, $90^\circ$ and $60^\circ$, respectively. Tx and Rx antennas numbers $N = 1$. Direct links are assumed to be blocked.}
    \label{fig:beam}
\end{figure}
We select a configuration where the BS, DL user, and UL user are located at  $30^\circ$, $90^\circ$ and $60^\circ$, respectively, and examine the normalized impinging and reflected beampatterns for DL and UL transmission. The impinging and reflected beampatterns for DL and UL transmission without structural scattering are defined by 
$P_{\mathsf{d}}^\mathsf{impinging}(\theta) =  |\mathbf{h}_\mathsf{d}^\top \mathbf{\Phi} \mathbf{a}(\theta)|^2$, $P_{\mathsf{d}}^\mathsf{reflected}(\theta) =  | \mathbf{a}^\top(\theta) \mathbf{\Phi} \mathbf{g}|^2$, $P_{\mathsf{u}}^\mathsf{impinging}(\theta) =  | \mathbf{g}^\top \mathbf{\Phi} \mathbf{a}(\theta)|^2$, and $P_{\mathsf{u}}^\mathsf{reflected}(\theta) =  | \mathbf{a}^\top(\theta) \mathbf{\Phi} \mathbf{h}_\mathsf{u}|^2$,
% \begin{subequations}
% \label{eq:bp}
% \begin{align}
%     P_{\mathsf{d}}^\mathsf{impinging}(\theta) &=  |\mathbf{h}_\mathsf{d}^\top \mathbf{\Phi} \mathbf{a}(\theta)|^2,    \\
%     P_{\mathsf{d}}^\mathsf{reflected}(\theta) &=  | \mathbf{a}^\top(\theta) \mathbf{\Phi} \mathbf{g}|^2, \\
%     P_{\mathsf{u}}^\mathsf{impinging}(\theta) &=  | \mathbf{g}^\top \mathbf{\Phi} \mathbf{a}(\theta)|^2, \\
%     P_{\mathsf{u}}^\mathsf{reflected}(\theta) &=  | \mathbf{a}^\top(\theta) \mathbf{\Phi} \mathbf{h}_\mathsf{u}|^2,
% \end{align}
% \end{subequations}
where $\mathbf{a} = \frac{1}{\sqrt{N} } [1, e^{\jmath \pi \cos(\theta)}, \cdots, e^{\jmath \pi (N-1) \cos(\theta)}]^\top \in \mathbb{C}^{N \times 1}$ is the steering vector, and $\theta \in [0, 180^\circ]$.
% Additionally, the impinging and reflected beampatterns with structural scattering can be calculated by changing $\mathbf{\Phi}$ to $(\mathbf{\Phi} - \mathbf{I})$ in all expressions \eqref{eq:bp}.
Additionally, the impinging and reflected beampatterns with structural scattering can be calculated by changing $\mathbf{\Phi}$ to $(\mathbf{\Phi} - \mathbf{I})$ in all expressions.
To plot the figure, we normalize the beampattern power and keep the relative difference of the beams, \eg $P_{\mathsf{d}}^\mathsf{impinging}(\theta)/P_{\mathsf{beam, max}}$, $P_{\mathsf{beam, max}} \triangleq \! \! \max \Big \{ \! \! \max\{ \! P_{\mathsf{d}}^\mathsf{impinging}\! (\theta) \! \}\!, \max\{ \! P_{\mathsf{d}}^\mathsf{reflected}\!(\theta) \! \}\!,\max\{ \! P_{\mathsf{u}}^\mathsf{impinging} \!(\theta) \! \}\!, \\ \max\{ P_{\mathsf{u}}^\mathsf{reflected}(\theta) \} \Big \}$.

% $$ \frac{ P_{\mathsf{d}}^\mathsf{impinging}(\theta)}
% {\max \{ \max\{ P_{\mathsf{d}}^\mathsf{reflected}(\theta) \}, \max\{ P_{\mathsf{u}}^\mathsf{impinging}(\theta) \},\\ \max\{ P_{\mathsf{d}}^\mathsf{impinging}(\theta) \}, \max\{ P_{\mathsf{u}}^\mathsf{reflected}(\theta) \} \}}.$$

Specifically, the DL impinging beampattern should probe towards the BS direction to support the transmission from the BS to the RIS. The DL reflected beampattern should probe at the DL user direction to support the transmission from the RIS to the DL user. Similarly, the UL impinging beampattern should probe at the UL user direction, and the UL reflected beampattern should probe at the BS direction. 
Starting from the beampattern performances without structural scattering, as shown in \fig{fig:beam} (i), all four beams probe towards their expected directions. In contrast, the DL impinging and reflected beams of both reciprocal BD-RIS and D-RIS are not directed at the required directions. Therefore, as shown in \fig{fig:angle} (b) and (e), the DL sum-rates of both reciprocal BD-RIS and D-RIS are significantly lower than that of the non-reciprocal BD-RIS. As a result, the total DL and UL sum-rate of non-reciprocal BD-RIS is higher than the sum-rates of other two types of RIS. Comparing the beampatterns with and without structural scattering, as shown in \fig{fig:beam} (i) and (ii) respectively, we observe that the structural scattering forces the DL reflected beam and UL impinging beam to probe at $150^\circ$, as previously explained specular reflection condition. In conclusion, the non-reciprocal BD-RIS effectively supports simultaneous DL and UL transmission in FD communications. In contrast, reciprocal BD-RIS and D-RIS do not adequately support DL and UL transmissions in FD mode.

\vspace{-10pt}
\subsection{Sum-rate Regions}
\label{ssec:sumrate_region}
\begin{figure}[t]
    \centering
    \includegraphics[width = 0.3\textwidth]{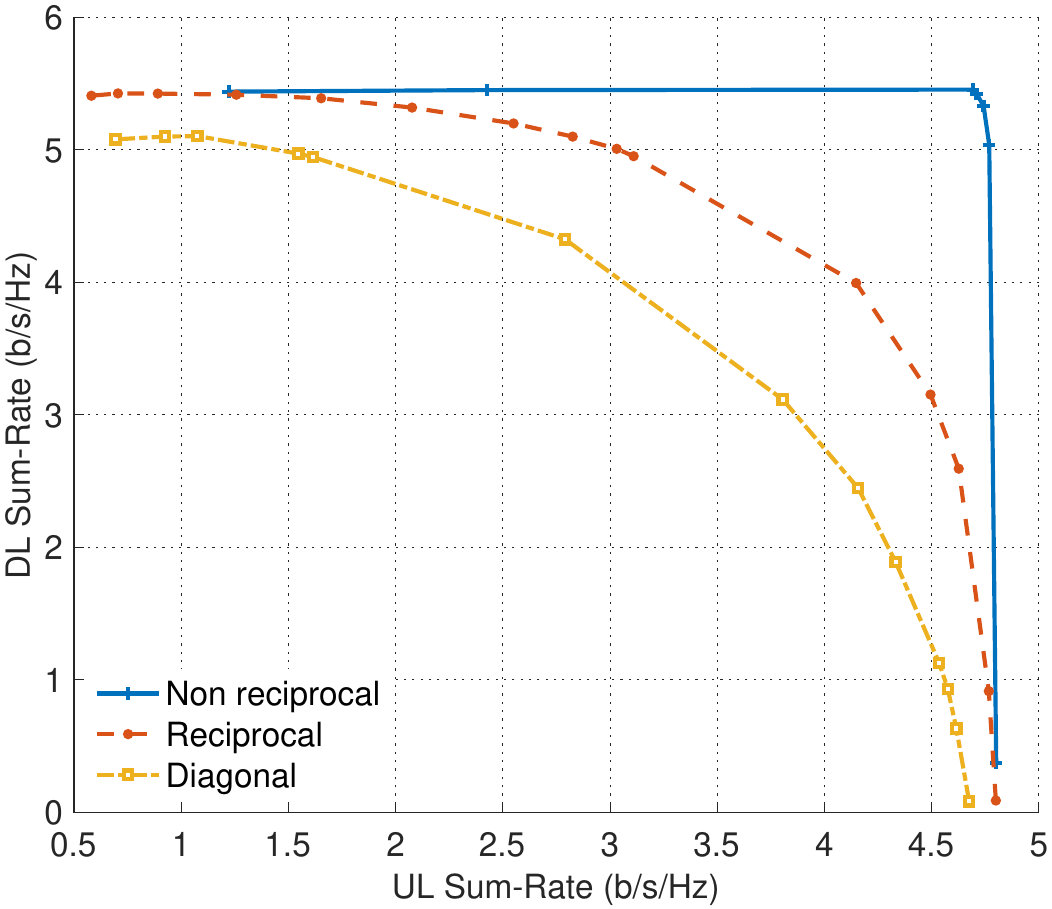}
    \centering
    \caption{The sum-rate regions for non-reciprocal BD-RIS, reciprocal BD-RIS, and D-RIS with structural scattering and BS location at $30^\circ$, DL user location at $90^\circ$ and UL user location at $60^\circ$, Tx and Rx antennas numbers $N = 1$. Direct links are assumed to be blocked.}
    \label{fig:rate_region}
\end{figure}

To understand the sum-rate regions of the three types of RISs, we vary the priority parameter $\alpha_\mathsf{d}$ between DL and UL transmission as in \eqref{eq:op1} from $\alpha_\mathsf{d} = 1$ (\ie only DL transmission) to $\alpha_\mathsf{d} = 0$ (\ie only UL transmission). Meanwhile, the locations of BS, DL user, and UL user are fixed at $30^\circ$, $90^\circ$ and $60^\circ$. {We assume the direct links are blocked.} As shown in \fig{fig:rate_region}, the sum-rate region of the non-reciprocal BD-RIS is larger compared to that of reciprocal BD-RIS and D-RIS when supporting both DL and UL transmissions in FD communications. Additionally, when only DL or UL transmission is required, the non-reciprocal BD-RIS achieves the same sum-rate performance as the reciprocal BD-RIS, while still outperforming the D-RIS. This demonstrates the distinct benefits of the non-reciprocal BD-RIS in FD scenarios.
\vspace{-10pt}
\subsection{DL and UL Sum-rates of Group-connected BD-RIS}
\begin{figure}[t]
    \centering
    \includegraphics[width = 0.3\textwidth]{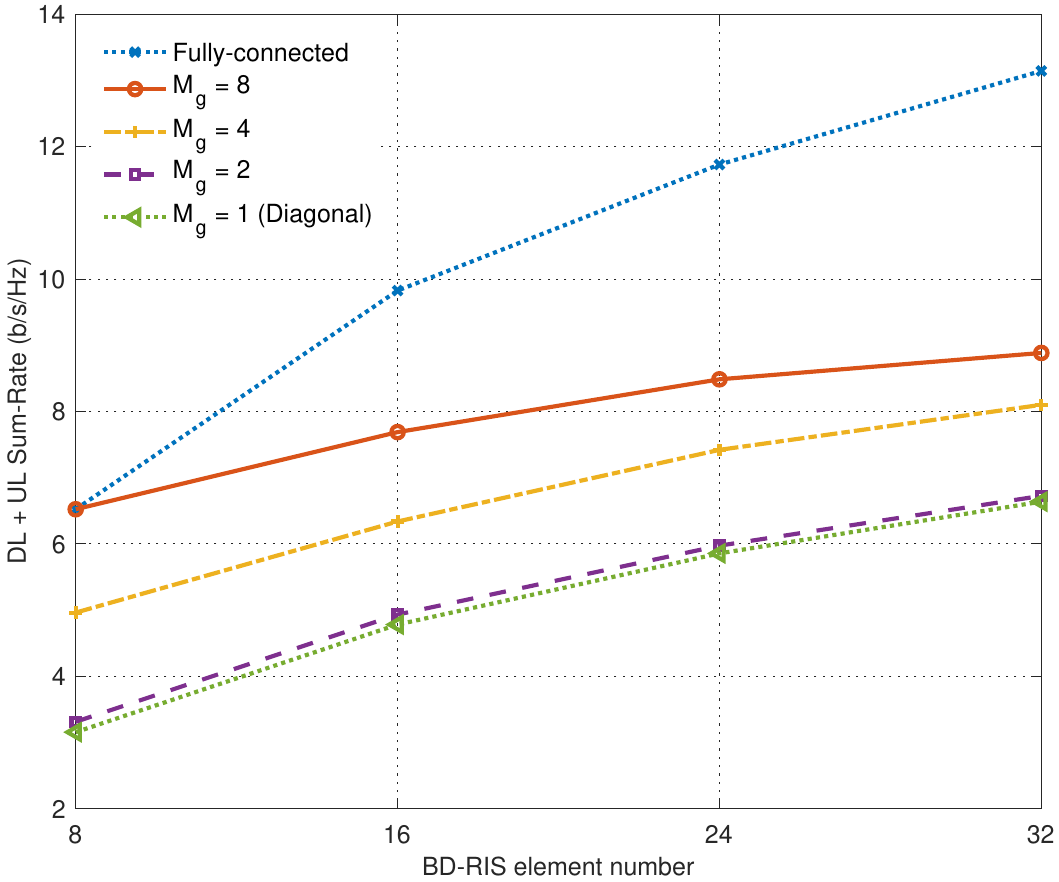}
    \centering
    \caption{The DL and UL sum-rates of non-reciprocal BD-RIS with structural scattering and different group size $M_g$. Direct links are assumed to be blocked. The locations are BS at $30^\circ$, DL user location at $90^\circ$ and UL user location at $60^\circ$. Tx and Rx antennas numbers $N = 1$.}
    \label{fig:group}
\end{figure}
\begin{figure}[t]
    \centering
    \includegraphics[width = 0.3\textwidth]{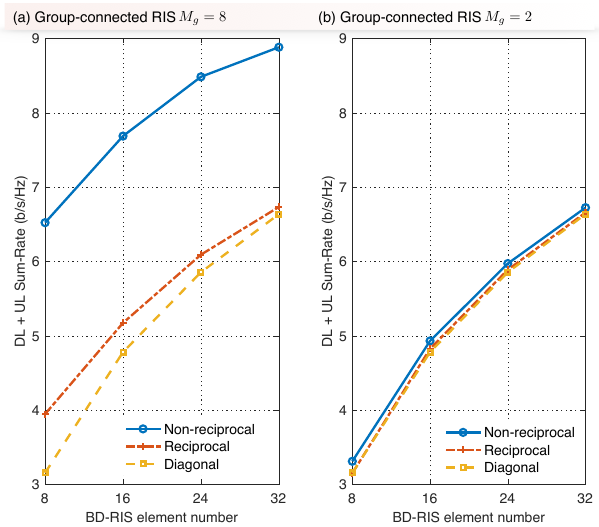}
    \centering
    \caption{{(a) The DL and UL sum-rates of non-reciprocal BD-RIS, reciprocal BD-RIS and D-RIS with fixed group size $M_g = 8$ and (b) $M_g = 2$. The locations are BS at $30^\circ$, DL user location at $90^\circ$ and UL user location at $60^\circ$. Tx and Rx antennas numbers $N = 1$. Direct links are assumed to be blocked.}}
    \label{fig:group_8_2}
\end{figure}
To understand the sum-rate performances of non-reciprocal BD-RIS, reciprocal BD-RIS and D-RIS with group-connected architecture, we maintain the same simulation conditions but vary the group size within the RISs. The structural scattering is captured in the group-connected case. {We also assume the direct links are blocked.} As shown in \fig{fig:group}, a larger group size leads to higher DL and UL sum-rates for the non-reciprocal BD-RIS. The lower bound is obtained with the D-RIS, which is also the single-connected RIS (\ie $M_g=1$). The upper bound is achieved by the fully-connected  non-reciprocal BD-RIS. \fig{fig:group_8_2} demonstrate that the non-reciprocal BD-RIS consistently achieves the highest DL and UL sum-rates compared to the other two types of RISs with a group-connected structure in the FD scenario. In addition, comparing \fig{fig:group} (a) and (b), it is observed that non-reciprocal BD-RIS achieves larger gain when group size increases.
\vspace{-10pt}
\subsection{The Impact of SI Power}
\begin{figure}[t]
    \centering
    \includegraphics[width = 0.3\textwidth]{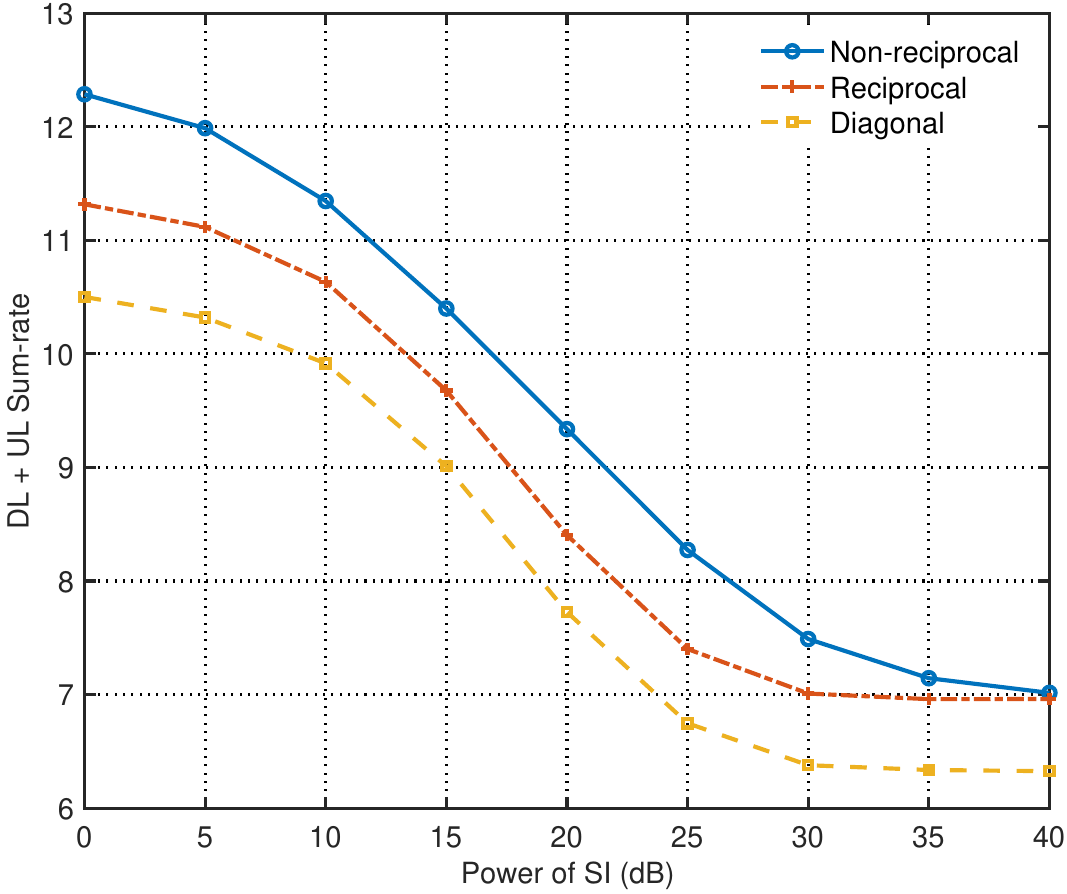}
    \centering
    \caption{{The DL and UL sum-rates of non-reciprocal BD-RIS, reciprocal BD-RIS and D-RIS with structural scattering versus the SI power. The locations are BS at $30^\circ$, DL user location at $150^\circ$ and UL user location at $75^\circ$. Tx and Rx antennas numbers $N = 1$ and the number of RIS element is $M=16$. Direct links are assumed to be blocked.}}
    \label{fig:si}
\end{figure}
We analyze the impact of SI power on the DL and UL sum-rates of the three RISs, with the BS, DL user, and UL user fixed at $30^\circ$, $150^\circ$, and $75^\circ$, respectively. The Tx/Rx antennas are $N = 1$, RIS elements $M = 16$, and direct links are blocked. In \fig{fig:si}, the sum-rates decrease with increasing SI power due to higher interference in the UL SINR \eqref{eq:sinr_dlandul}. Non-reciprocal BD-RIS achieves the highest sum-rates, but its gain over others decreases as SI power rises. This is because, at high SI levels, \eg $40$ dB, the UL rate becomes negligible, making the sum-rate DL-dominated. This is also the reason why the curves become flat, as they represent DL rates. As explained in \ref{ssec:sumrate_region}, when one-direction transmission is considered, non-reciprocal has no benefits, while reciprocal BD-RIS outperforms D-RIS due to its fully-connected structure enabling control of both amplitude and phase.
\vspace{-10pt}
\subsection{{MU Sum-rates Regions}}
\begin{figure}[t]
    \centering
    \includegraphics[width = 0.49\textwidth]{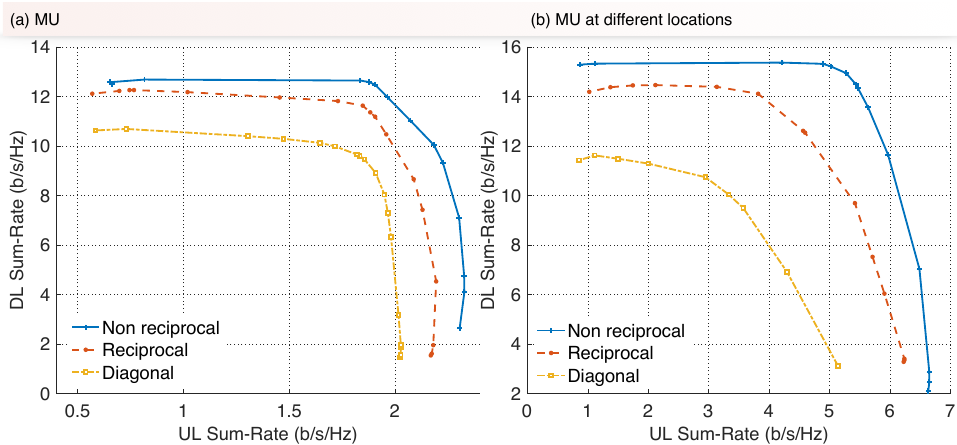}
    \centering
    \caption{The sum-rate regions for non-reciprocal BD-RIS, reciprocal BD-RIS, and D-RIS with structural scattering and RIS element $M=16$, and fixed BS location at $30^\circ$. Direct links are assumed to be blocked. (a) The $K=2$ DL users are aligned at $90^\circ$, and $I=2$ UL users are aligned at $60^\circ$. (b) The $K=2$ DL users are located at $90^\circ$ and $120^\circ$, and $I=2$ UL users are located at $60^\circ$ and $75^\circ$. Tx and Rx antennas numbers are $N = 2$.}
    \label{fig:rate_region_MU}
\end{figure}
We also examine the sum-rate regions of the MU cases. In the first case, we assume that $K=2$ DL users are aligned at $90^\circ$, and $I=2$ UL users are aligned at $60^\circ$. Additionally, we consider a non-aligned scenario where $K=2$ DL users are located at $90^\circ$ and $120^\circ$, and $I=2$ UL users are located at $60^\circ$ and $75^\circ$, respectively. {The direct links are assumed to be blocked.} The BS has $N = 2$ transmit and receive antennas. The weight $\alpha_\mathsf{d}$ between DL and UL communications is adjusted to prioritize each function. In both two cases, as expected, the sum-rate regions of non-reciprocal BD-RIS are larger than the ones of reciprocal BD-RIS and D-RIS. Furthermore, the sum-rate regions for multiple DL and UL users at different locations are larger than those for aligned DL and UL users. This increase is due to the reduced channel correlation and, consequently, the higher diversity in the non-aligned scenario.

\section{Conclusion}
\label{sec:con}
{In this paper, we have investigated the use of non-reciprocal, reciprocal BD-RISs and D-RIS in FD communication systems considering both direct and reflected links.} Our system model captures the  loop interference at the BS, UL-to-DL user interference, and structural scattering.  This problem has been formulated as a non-convex optimization problem, where we aim to maximize the DL and UL sum-rates by designing the precoder and combiner in the BS, and scattering matrix. An iterative algorithm based on BCD framework and PDD has been proposed to address the problem. Numerical results have demonstrated the superiority of the non-reciprocal BD-RIS compared to reciprocal BD-RIS and D-RIS in FD communications, as the non-reciprocal BD-RIS support DL and UL users in different directions in single and multiple user cases. The gain of non-reciprocal BD-RIS over other two kinds of RIS increases as the number of RIS elements and group size increase, due to the higher flexibility achieved by the reconfigurable elements.  {This work paves the way for further studies on non-reciprocal BD-RIS and enables future work on non-reciprocal hybrid/multi-sector BD-RIS in FD.}

\bibliographystyle{IEEEtran_url}
\bibliography{IEEEabrv,references}

\end{document}